\documentclass[lettersize,journal]{IEEEtran}
\usepackage{mathrsfs}
\usepackage{booktabs}
\usepackage{microtype}
\usepackage{amsmath,amsthm,amssymb}
\usepackage{float}
\usepackage{microtype}
\usepackage{graphicx,epsfig,url}
\usepackage[linesnumbered,boxed]{algorithm2e}
\usepackage{subfig}
\usepackage{cite}
\usepackage[usenames, dvipsnames]{color}  
\usepackage{mathtools,mathrsfs,bm}
\usepackage{multirow}
\usepackage{adjustbox}
\usepackage{tabularx,booktabs}
\newcolumntype{Y}{>{\centering\arraybackslash}X}
\usepackage[pagebackref,breaklinks,colorlinks]{hyperref}

\usepackage[capitalize]{cleveref}
\crefname{section}{Sec.}{Secs.}
\Crefname{section}{Section}{Sections}
\Crefname{table}{Table}{Tables}
\crefname{table}{Tab.}{Tabs.}

\def\i{\boldsymbol{i}}
\def\j{\boldsymbol{j}}

\def\k{\boldsymbol{k}}
\def\btau{\boldsymbol{\tau}}
\def\bomega{\boldsymbol{\omega}}

\def\P{\boldsymbol{P}}

\def\N{\boldsymbol{N}}

\def\Re{\mathbb{R}}
\def\F{\mathbf{F}}

\def\W{\mathbf{W}}
\def\P{\mathbf{P}}
\def\Pr{\mathscr{P}}

\def\Q{\mathbf{Q}}
\def\C{\mathscr{C}}
\def\Y{\mathbf{Y}}
\def\X{\mathbf{X}}
\def\Z{\mathbf{Z}}
\def\E{\mathbf{E}}
\def\B{\mathbf{B}}
\def\S{\mathbf{S}}

\def\C{\mathbf{C}}
\def\N{\mathbf{N}}
\def\R{\mathbf{R}}
\def\V{\mathbf{V}}
\def\U{\mathbf{U}}
\def\W{\mathbf{W}}
\def\I{\mathbf{I}}

\def\D{\mathbf{D}}
\def\Z{\mathbf{Z}}

\newcommand{\inner}[2]{\left\langle#1,#2\right\rangle}

\def\BLambda{\boldsymbol{\Lambda}}
\def\BSigma{\boldsymbol{\Sigma}}
\newtheorem{theorem}{Theorem}

\newcommand*\xbar[1]{%
   \hbox{%
     \vbox{%
       \hrule height 0.5pt 
       \kern0.5ex
       \hbox{%
         \kern-0.1em
         \ensuremath{#1}%
         \kern-0.1em
       }%
     }%
   }%
}

\begin{document}


\title{Guided Nonlocal Patch Regularization and Efficient Filtering-Based Inversion for Multiband Fusion}

\author{Unni V. S.,~\IEEEmembership{Student~Member,~IEEE}, Pravin Nair,~\IEEEmembership{Student~Member,~IEEE} and Kunal N. Chaudhury,~\IEEEmembership{Senior~Member,~IEEE}

\thanks{The work of K.~N.~Chaudhury was supported by  grant CRG/2020/000527  from SERB, Government of India and ISRO-IISc Space Technology Cell grant ISTC/EEE/KNC/440}
\thanks{Unni V.~S., Pravin Nair, and K.~N.~Chaudhury are with the Department of Electrical Engineering, Indian Institute of Science, Bengaluru 560012, India. Email: unniv@iisc.ac.in, pravinn@iisc.ac.in, kunal@iisc.ac.in.}
}

\maketitle

\begin{abstract}
In multiband fusion, an image with a high spatial and low spectral resolution is combined with an image with a low spatial but high spectral resolution to produce a single multiband image having high spatial and spectral resolutions. This comes up in remote sensing applications such as pansharpening~(MS+PAN), hyperspectral sharpening~(HS+PAN), and HS-MS fusion~(HS+MS). Remote sensing images are textured and have repetitive structures. Motivated by nonlocal patch-based methods for image restoration, we propose a convex regularizer that (i) takes into account long-distance correlations, (ii) penalizes patch variation, which is more effective than pixel variation for capturing texture information, and (iii) uses the higher spatial resolution image as a guide image for weight computation. We come up with an efficient ADMM algorithm for optimizing the regularizer along with a standard least-squares loss function derived from the imaging model. The novelty of our algorithm is that by expressing patch variation as filtering operations and by judiciously splitting the original variables and introducing latent variables, we are able to solve the ADMM subproblems efficiently using FFT-based convolution and soft-thresholding. As far as the reconstruction quality is concerned, our method is shown to outperform state-of-the-art variational and deep learning techniques.
\end {abstract}

\begin{keywords}
Hyperspectral imaging, image fusion, patch regularization, filtering, optimization.
\end{keywords}
\section{Introduction}
\label{intro}

With the rapid advent of satellite imaging, remote sensing images are now widely used for object classification and recognition \cite{yokoya2017hyperspectral}, tracking \cite{UHV2016}, environmental monitoring \cite{PDBJK2011}, etc. Multiband imaging sensors acquire information as a band of two-dimensional images, where each image captures a narrow wavelength of the electromagnetic spectrum. The precision of these sensors is determined by their spectral and spatial resolutions, i.e., how densely the spectral and spatial domains are sampled. However, existing sensors have a fundamental tradeoff between spatial and spectral resolutions \cite{yokoya2017hyperspectral,simoes2015convex}. For example, hyperspectral (HS) images have hundreds of bands but low spatial resolution, while multispectral (MS) images have fewer bands but higher spatial resolution. An extreme case is a panchromatic (PAN) image which has a single band, but its spatial resolution is much higher than HS and MS images.  Reconstruction of high-resolution multiband images by fusing images with complementary spatial and spectral resolutions is thus of great interest \cite{yokoya2017hyperspectral}. 


\subsection{Variational models}

Many existing techniques for multiband image fusion are based on component substitution   \cite{ihs,gs,rvs,survey,hpf,aiazzi2007improving,yokoya2017hyperspectral}, multiresolution analysis  \cite{mtfglp,mtfglppp,mtfglphpm,loncan2015hyperspectral} or spectral unmixing \cite{yokoya2012coupled,iccv15}. 
More recent model-based variational methods have been shown to yield state-of-the-art results \cite{thomas2008synthesis,simoes2015convex,loncan2015hyperspectral, yokoya2017hyperspectral}. The standard image-formation model used in these works is
\begin{equation}
\label{model}
\Y_\ell = \S\B\Z + \N_\ell \quad \text{and} \quad \Y_h = \Z\R + \N_h,
\end{equation}
where $\Y_\ell \in \Re^{n_\ell \times L_\ell}$ is the observed low spatial-high spectral resolution image with $L_\ell$ bands and $n_\ell$ pixels, $\Y_h \in \Re^{n_h\times L_h}$ is the observed high spatial-low spectral resolution image with $L_h$ bands, and $\Z \in \Re^{n_h \times L_\ell}$ is the ground-truth image having high spatial and spectral resolutions. Typically, $L_h \ll L_\ell$  and $n_\ell \ll n_h$. In this work, we assume that the observed images $\Y_\ell$ and $\Y_h$ are preprocessed and aligned \cite{wei2015bayesian,wei2015hyperspectral,wei2015fast}. The intrinsic parameters in \eqref{model} are the blur operator $\B \in \Re^{n_h \times n_h}$, the decimation operator $\S \in \Re^{n_\ell \times n_h}$, and  the spectral response operator $\R \in \Re^{L_\ell \times L_h }$ \cite{hardie2004map,molina2008variational}; we assume that these parameters are known. $\N_\ell$ and $\N_h$ are white Gaussian noise. 
Though the variable $\Z$ lies in a high-dimensional space ($n_h L_\ell$ unknowns), it has a low rank due to the correlation between  different bands. This observation is used to  cut down the number of variables via the low-rank representation $\Z=\X\E$, where $\E \in \Re^{L_s \times L_\ell}$ is a fixed matrix (obtained using SVD or vertex component analysis \cite{nascimento2005vertex} of $\Y_\ell$) and $\X \in \Re^{n_h \times L_s}$ is the low-dimensional variable having fewer unknowns \cite{simoes2015convex,wei2016r,wei2015fast}.
Combining this with \eqref{model}, the image fusion task can be posed as a variational problem \cite{simoes2015convex}:
\begin{equation}
\label{mainoptim}
\min_{\X} \ \ \ell(\X)  + \lambda_2 \phi(\X), 
\end{equation}
where 
\begin{equation*}
\ell(\X) = \frac{1}{2}\|\Y_\ell - \S\B\X\E\|^2 +   \frac{\lambda_1}{2} \|\Y_h - \X\E\R\|^2
\end{equation*}
is a model-based loss function with $\|\cdot\|$ being the Frobenius norm, $\phi: \Re^{n_h \times L_s} \to \Re$ is a regularizer, and $\lambda_1, \lambda_2 > 0$ are tunable parameters that are used to balance the loss and regularization terms. 

The regularizer $\phi(\X)$ plays a crucial role in determining the fusion quality and the complexity of the associated  reconstruction algorithm. Different regularizers have been proposed over the years, many of which produce state-of-the-art results for multiband fusion. The authors in \cite{simoes2015convex} introduced a multiband extension of total variation called vector total variation (VTV), and solved the associated optimization problem using the alternating directions method of multipliers (ADMM). Sparse dictionary-based regularization was proposed in \cite{wei2015hyperspectral}, where the dictionary and the corresponding sparse coefficients are learnt from the observed images; the associated optimization is solved using ADMM. For pansharpening, the best performing results are obtained using a variant of VTV where the gradients of the reconstructed image are matched with the observed MS image \cite{chen2014image,fu2019variational}; the optimization is performed 
using fast iterative shrinkage. The authors in \cite{wei2015bayesian} proposed a Bayesian regularizer where the associated optimization is performed using ADMM and block coordinate descent. 

\subsection{Nonlocal regularization}

Nonlocal regularization models have been shown to outperform pixel-based regularization for inverse problems such as superresolution, inpainting, and compressed sensing \cite{MBPSZ2009,nl2,peyre,ZW2011}. 
The challenge with nonlocal regularizers is to come up with efficient iterative algorithms. 
Split Bregman is used in \cite{nl2}, alternating minimization in \cite{peyre,MBPSZ2009}, and half-quadratic splitting in \cite{ZW2011}. However, except for \cite{nl2}, the regularizer in \cite{MBPSZ2009,peyre,ZW2011} are nonconvex and the algorithms do not come with any convergence guarantee. The regularizer in \cite{nl2} uses pixels to measure nonlocal variations.
As explained in Section \ref{solver}, computational challenges arise in extending the split Bregman optimizer (which is similar to ADMM) in \cite{nl2} to handle patch-based variations.

Nonlocal patch-based methods have been around for quite some time. In \cite{liu2016nonlocal}, the authors introduced a variant of nonlocal TV for  denoising, inpainting, and compressive sensing. The optimization problem is solved using ADMM. Nonlocal TV is used for tomographic reconstruction in \cite{lou2010image}. In \cite{zhang2019hyperspectral}, a combination of nonlocal low-rank tensor decomposition and TV  regularization is used for hyperspectral denoising; the model  is solved using ADMM. Nonlocal low-rank regularization is used for compressive sensing in \cite{dong2014compressive}. The authors in \cite{liu2017image} used nonlocal intra- and inter-patch correlations for regularization purpose; the model  is solved using IRLS algorithm. In \cite{lu2019admm},  each patch is modeled as a sparse linear combination from an over-complete dictionary and this is used for the regularization of general image restoration tasks using ADMM. In \cite{yao2019nonconvex}, the problem of spectral unmixing is solved using nonlocal TV within the ADMM framework.

Nonlocal regularization techniques have also been used for image fusion. The authors in \cite{duran2014nonlocal} used weighted nonlocal pixel regularization for pansharpening and gradient descent was used to solve the optimization problem.
In \cite{duran2017survey}, the authors used the regularizer in \cite{duran2014nonlocal} along with a ``radiometric constraint'' that is used to preserve the radiometric ratio between the panchromatic and the spectral bands. The resulting optimization is also solved using gradient descent.
The authors in \cite{mifdal2021variational} used nonlocal pixel regularization and a primal-dual algorithm for hyperspectral fusion. Apart from the data-fidelity terms corresponding to $\Y_\ell$ and $\Y_h$, the authors also used radiometric constraints.
 In \cite{duran2018restoration}, the authors used weighting and nonlocal variations similar to \cite{duran2014nonlocal}. However, instead of  the $\ell_2$ norm on the pixel gradients as in \cite{duran2014nonlocal}, the authors used the $\ell_1$ norm. The optimization is performed using a primal-dual algorithm.




\subsection{Deep learning models}

Recently, deep learning methods have been shown to produce promising results for multiband fusion \cite{huang2015new,masi2016pansharpening,wei2017boosting,palsson2017multispectral,shao2018remote,xie2019multispectral,zhang2021image,zhang2021sdnet,wu2021dynamic,fang2020cross,wang2021enhanced,jin2021bam,xu2020u2fusion,fu2020deep}. These methods use a deep neural network to learn the end-to-end functional relationship between the observed  and ground truth images.
Although these methods have excellent performance, sufficient ground-truth is required for training a deep neural network, and this is difficult for remote sensing applications. Limited data ultimately affects the generalization capacity of the trained network \cite{friedman2017elements,fu2019variational}. 
Last, but not the least, the network is trained to work with a fixed forward model and cannot directly be used for other models. It is thus not surprising that state-of-the-art techniques for multiband fusion are based on variational models \cite{fu2019variational}.

\subsection{Contributions}

In this work, we adopt the variational model in \eqref{mainoptim}. The novelty is the form of the nonlocal regularizer $\phi(\X)$ and an efficient filtering-based solution of the associated optimization problem.

 \textbf{(i) Regularization model}: Motivated by prior work on nonlocal regularization, we introduce a regularizer for remote sensing images that can exploit self-similar structures such as textures using image patches. However, unlike existing nonlocal regularizers \cite{MBPSZ2009,peyre,ZW2011}, our regularizer is convex. We note that the convex nonlocal regularizer in \cite{nl2} is designed to capture pixel-based variations. In contrast, we use patches for penalizing variations jointly across all bands. Moreover, unlike VTV and local regularizers \cite{lefkimmiatis2015structure, bredies2010total, chan2000high}, where variations along just horizontal and vertical directions are considered, we use a wider search window in terms of nonlocality and directionality. Finally, since in fusion we have a high resolution image of the same scene, we use this as a guide for computing weights, which are incorporated in the regularizer. 
 
 \textbf{(ii) Optimization algorithm}: In Section \ref{conv}, we point out that, owing to the use of patch variations, a straightforward ADMM solution of \eqref{mainoptim} leads to computational issues. Nevertheless, we show that by expressing patch variation using convolutions, and by judiciously splitting the original variables and introducing new latent variables, we can efficiently solve the ADMM subproblems using FFT-based convolutions and soft-thresholding. Importantly, since our regularizer is convex, we are able to guarantee convergence of the ADMM iterations to a global optimum.

\section{Regularization model}

\begin{figure*}
\centering
\subfloat[Clean.]{\includegraphics[width=0.16\linewidth]{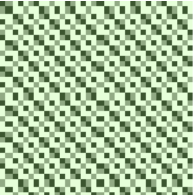}} \hspace{0.05mm}
\subfloat[Mask.]{\includegraphics[width=0.16\linewidth]{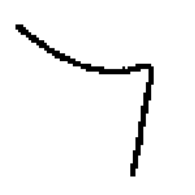}} \hspace{0.05mm}
\subfloat[Observed.]{\includegraphics[width=0.16\linewidth]{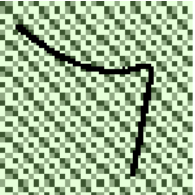}} \hspace{0.05mm}
\subfloat[Local.]{\includegraphics[width=0.16\linewidth]{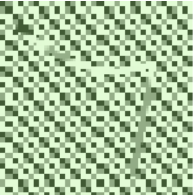}} \hspace{0.05mm}
\subfloat[Unguided NLPR.]{\includegraphics[width=0.16\linewidth]{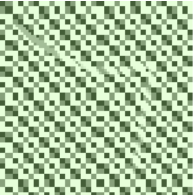}} \hspace{0.05mm}
\subfloat[Guided NLPR.]{\includegraphics[width=0.16\linewidth]{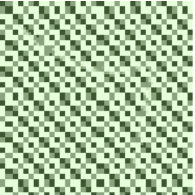}}  \hspace{0.05mm}
\\~\\
\caption{Inpainting of a synthetic RGB image. The weights $\omega_{\i\btau}$ are set to $1$ for (e) and are computed from a grayscale guide for (f). Notice that the local pixel-based method  \cite{bresson2008fast} performs poorly around the missing pixels and is unable to restore the texture patterns there. Also notice that we can further enhance the restoration quality  by using guide-based weighting.}
\label{toy_example}
\end{figure*}

\begin{figure*}
\centering
\subfloat[Clean \cite{CAVE_0293}.]{\includegraphics[width=0.16\linewidth]{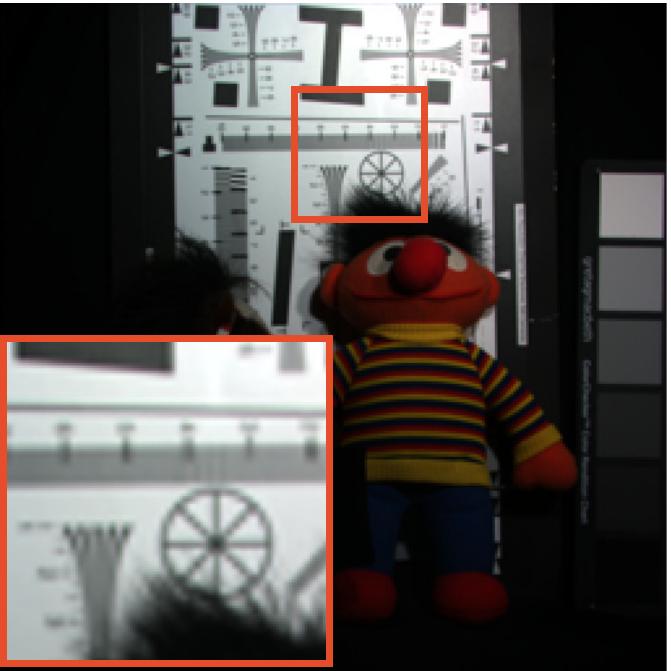}} \hspace{0.05mm}
\subfloat[Noisy.]{\includegraphics[width=0.16\linewidth]{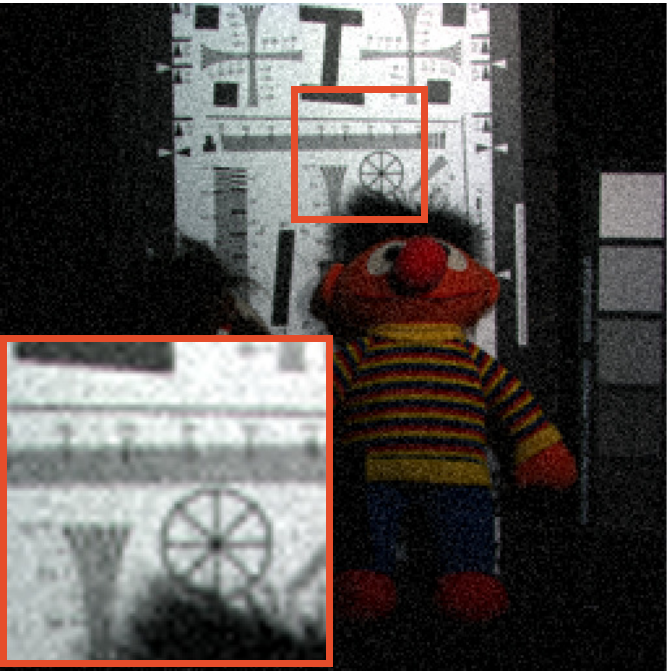}} \hspace{0.05mm}
\subfloat[VTV.]{\includegraphics[width=0.16\linewidth]{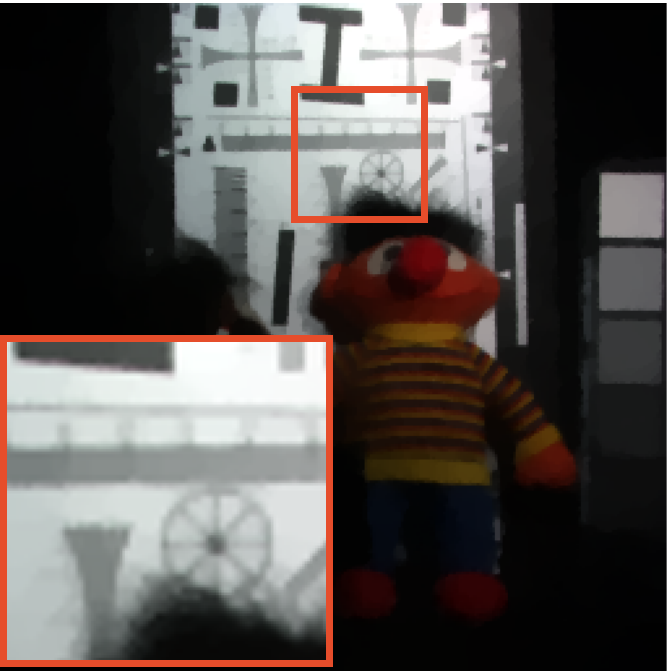}} \hspace{0.05mm}
\subfloat[BM4D.]{\includegraphics[width=0.16\linewidth]{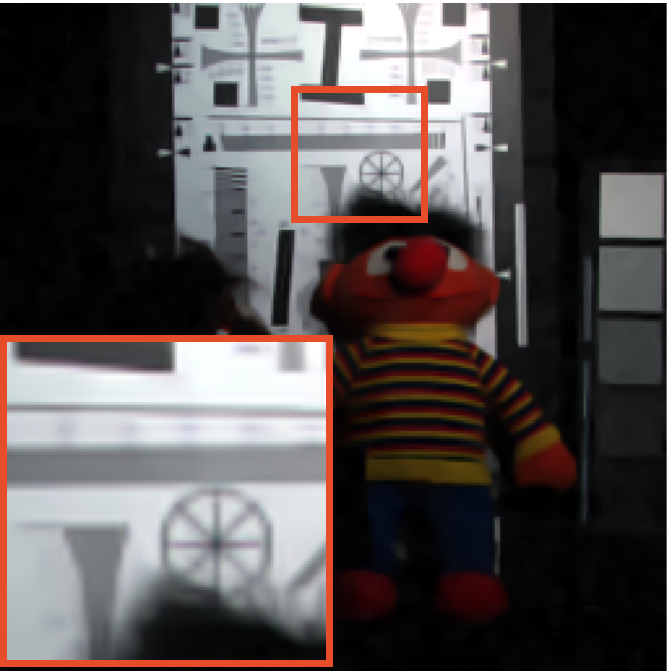}}  \hspace{0.05mm}
\subfloat[Unguided NLPR.]{\includegraphics[width=0.16\linewidth]{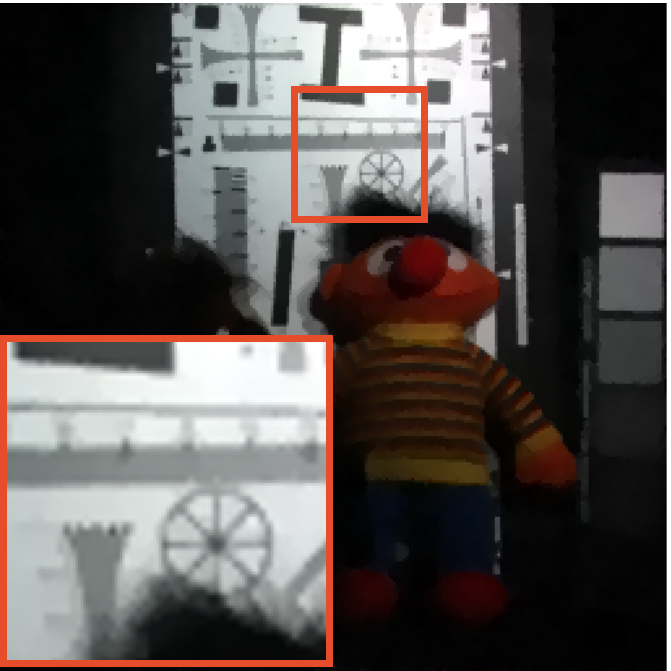}}  \hspace{0.05mm}
\subfloat[Guided NLPR.]{\includegraphics[width=0.16\linewidth]{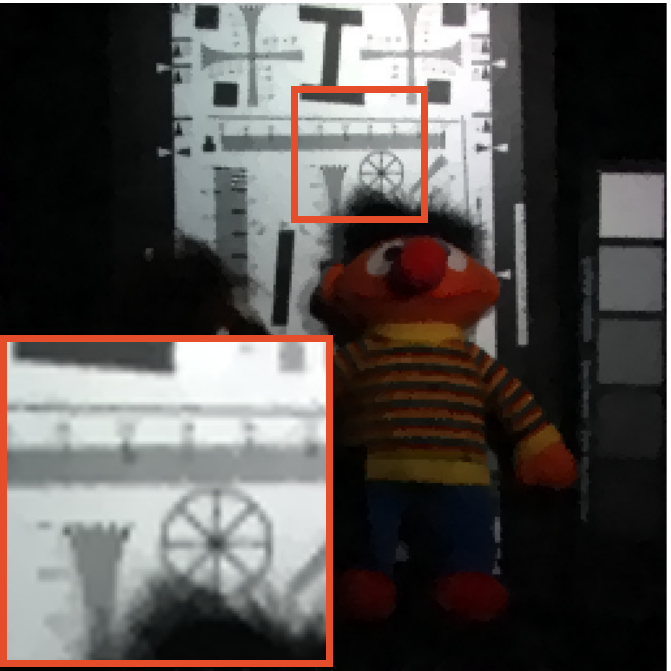}}  \hspace{0.05mm}
\\~\\
\caption{Multiband denoising results. Only three bands are shown for visualization. The weights $\omega_{\i\btau}$ are set to $1$ for (e) and the weights for (f) are computed from a RGB representation of the multiband image. The (PSNR, SSIM) values are as follows: (b)  ($28.12$, $0.5548$), (c)  ($32.57$, $0.9107$), (d)  ($34.07$, $0.9129$), (e)  ($33.60$, $0.9156$), (f) ($34.21$, $0.9217$). See the zoomed regions for a better comparison.}
\label{denoising_results}
\end{figure*}


Local pixel-based methods  have powerful regularization capability \cite{lefkimmiatis2015structure,bredies2010total,chan2000high}. However, in many applications, their suitability is limited by the type of signal they can promote. For example, total variation (TV) can preserve strong edges but promotes piecewise-constant solutions, and is hence less effective for capturing textures and geometrical patterns. In particular, TV cannot satisfactorily capture repetitive nonlocal structures such as textures. It has been demonstrated that this problem can be addressed using nonlocal regularization \cite{nl2,nl3,peyre,GO2009,gilboa2007nonlocal}, i.e., by using image patches for comparison and by exploiting long-range similarities. A major bottleneck in using nonlocal regularization for remote sensing applications such as pansharpening and hyperspectral fusion is that the inversion process derived from such  models is computationally intensive. We address this problem  in the present work.

\subsection{Motivation}

Motivated by prior work \cite{MBPSZ2009,nl2,peyre,ZW2011}, we propose a nonlocal patch-based regularizer for multiband remote sensing images.
Unlike VTV \cite{bresson2008fast}, where variations along just horizontal and vertical directions are penalized, we use a wider search window both in terms of nonlocality and directionality.  
Another novelty of our proposal  is the use of patches for computing weights which are used in the regularizer. While the idea of weighting is not exactly new, the distinction is precisely in how the weighting is performed. For example, in \cite{nl2,peyre} the weights are derived from the reconstruction and this makes the regularizer and the overall optimization nonconvex. Indeed, unlike image denoising \cite{buades2005non}, a good surrogate of the ground-truth is not available for problems such as deblurring and inpainting  and coupling the weight computation with the reconstruction seems to be a natural option. 
However, for the fusion problem at hand, we have a high resolution image $\Y_h$ of the same scene as the reconstruction and we use this 

\subsection{Definition}

To formulate the proposed nonlocal patch regularizer (referred to as NLPR), we introduce some notations. We let $\Omega=[0,p-1] \times [0,q-1]$  denote the image domain (with periodic boundary conditions), where $n_h=pq$ is the total number of pixels. We let $X_c: \Omega \to \Re$ denote the $c$-th image band, i.e., the $c$-th column of matrix $\X$. 
This will shortly be used for defining the patch operator and will also come up later for defining convolutions in Section \ref{conv}. The reason we introduce $X_c$ is that it is difficult to define shifts and convolutions directly on the columns of $\X$ (representing different bands). 

Our regularizer uses patches from both the reconstruction $\X$ and the observed image $\Y_h$. We use circular shifts on $\Omega$ which goes with the circular convolution and the discrete Fourier transform that are used later. In particular, 
for $\i \in \Omega$ and $\btau \in \mathbb{Z}^2$, we define 
\begin{equation*}
\i-\btau=(i_1-\tau_1 \ \mathrm{mod} \,p, i_2-\tau_2 \ \mathrm{mod} \, q).
\end{equation*}
Note that $\i-\btau \in \Omega$ by definition.

Assume that the number of channels $L$ is either $L_s$ or $L_h$.  For $\i \in \Omega$, the patch-extraction operator $\Pr_{\i}: \Re^{n_h \times L} \to \Re^{(2K+1)^2L}$ is defined as follows. Let  $\X \in \Re^{n_h \times L}$. We first fix a one-dimensional ordering of the set 
\begin{equation*}
\Lambda = \Big\{ (\k,c) : \ \k \in [-K,K]^2, \, c \in [1, L]\Big\},
\end{equation*}
where $K$ is the patch length. With respect to this ordering, we consider the  intensity values $X_c(\i - \k), (\k,c) \in \Lambda,$ and represent it as a vector of length $(2K+1)^2L$. This is precisely $\Pr_{\i}(\X)$.

For given $\i \in \Omega$ and $\btau \in \mathbb{Z}^2$, we define the patch-difference operator $\Pr_{\i \btau}: \Re^{n_h \times L} \to \Re^{(2K+1)^2L}$ to be 
\begin{equation*}
\Pr_{\i \btau}(\X)=\Pr_{\i}(\X) - \Pr_{\i - \btau}(\X),
\end{equation*}
i.e., $\Pr_{\i \btau}(\X)$ is the difference of the patch vectors around $\i$ and $\i - \btau$, where $\btau$ corresponds to shifts over the search window $W=[-S,S]^2$ with $S$ being sufficiently large. We skip the description of $\Pr_{\i \btau}(\Y_h)$ which is defined similarly.

The proposed NLPR regularizer is defined as
\begin{equation}
\label{reg1}
\phi(\X) =  \sum_{\i \in \Omega} \sum_{\btau \in W} \omega_{\i\btau}  \Vert \Pr_{\i\btau}(\X) \Vert_1,  
\end{equation}
where
\begin{equation}
\label{wts}
\omega_{\i\btau} =  \exp\!\Big( -\frac{1}{h^2} \Vert \Pr_{\i\btau}(\Y_h) \Vert^2_{2}\Big),
\end{equation}
$h > 0$ is a smoothing parameter and  $\Vert \cdot \Vert_p$ is the $\ell_p$ norm.  The precise reason for using the weighted $\ell_1$ norm in \eqref{reg1} is that its proximal map has a closed form solution. Note that the weights in \eqref{wts} are similar to that used in nonlocal means \cite{buades2005non}.

We use the $\ell_1$ norm due to its better robustness to outliers compared to the $\ell_2$ norm; this is a well-known property of the $\ell_1$ norm \cite{duran2015novel}. In fact, this is also the choice of norm in recent works that have been shown to perform well \cite{mifdal2021variational,duran2018restoration}.

As a regularizer, $\phi(\X)$ measures the weighted patch variations within $\X$ over a nonlocal neighborhood, where the weights are computed from patches extracted from the observed high-resolution image $\Y_h$.
The proposed weighting ensures that if pixels $\i$ and $\i-\btau$ are relatively close in terms of the patch distance $\Vert \Pr_{\i\btau}(\Y_h) \Vert_2$, then a larger penalty is incurred if the corresponding patches in the reconstruction $\X$ are very different. An important technical point that will be useful later is that $\phi(\X)$ is a convex function of $\X$. However, it is not differentiable and we cannot use gradient-based algorithms.

\subsection{Regularization capacity}

We will show later in the section~\ref{experiments} that NLPR outperforms state-of-the-art fusion techniques and is better able to capture texture patterns. In fact, NLPR is generally more powerful than local pixel regularizers and the improvement is visually apparent for  applications such as inpainting and denoising. Two such results are shown in figures \ref{toy_example} and \ref{denoising_results}.
For denoising, we have used a RGB representation of the noisy multiband image as a guide. Since a reliable guide image is not available for inpainting, we have used a grayscale representation of the original image as a guide. We remark that such a guide is not available in practice for inpainting; the idea is simply to show how the restoration quality can be enhanced by using guide-based weighting. This is particularly apparent from the results in figure \ref{toy_example}. We have also compared with local pixel-based VTV \cite{bresson2008fast}. Notice that the restored texture is visibly better for guided NLPR compared to VTV and unguided NLPR. For the denoising result in figure \ref{denoising_results}, notice that tiny details are  smoothed out by VTV and BM4D \cite{BM4D} but NLPR is able to retain fine structures.

\section{Patch variation using convolutions}
\label{conv}

On plugging the proposed NLPR regularizer \eqref{reg1} into \eqref{mainoptim}, we obtain the optimization problem:
\begin{align}
\label{ouroptim}
\underset{\X}{\min} & \quad \frac{1}{2}{\|\Y_\ell-\S\B\X\E\|}^2+\frac{\lambda_1}{2}{\|\Y_h-\X\E\R\|}^2\! \nonumber \\
&+ \frac{\lambda_2}{2} \sum_{\i \in \Omega} \sum_{\btau \in W} \omega_{\i\btau} {\Vert\Pr_{\i\btau}(\X)\Vert}_1. 
\end{align}
We wish to come up with an efficient solver for \eqref{ouroptim}. This is an unconstrained nonsmooth convex optimization, where the objective function is not differentiable due to the presence of the $\ell_1$ norm. However, since the proximal map of the weighted $\ell_1$ norm has a closed-form solution, this can be solved using a proximal algorithm such as ADMM  \cite{parikh2014proximal,boyd2011distributed}. The advantage of ADMM is that the original variable can be split into latent variables so that the ensuing subproblems can be solved efficiently, often in closed-form.
For example, a straightforward splitting for \eqref{ouroptim} results in the following equivalent problem:
\begin{equation}
\begin{gathered}
\label{fusionoptim2}
\min \ \frac{1}{2}\|\Y_\ell - \S\P_1\E\|^2  + \frac{\lambda_1}{2}{\|\Y_h - \P_2\E\R\|}^2 \! + \!\frac{\lambda_2}{2} g(\Q)  \\
\mbox{s.t.} \quad \P_1=\B\X, \quad \P_2=\X, \\ 
\Pr_{\i\btau}(\X) = \Q_{\i\btau}  \quad  (\i \in \Omega, \btau \in W),
\end{gathered}
\end{equation}
where 
\begin{equation*}
g(\Q) = \sum_{\i \in \Omega} \sum_{\btau \in W} \omega_{\i\btau}  \Vert \Q_{\i\btau} \Vert_1.
\end{equation*}
The variables in this case are $\X,\P_1,\P_2$ and $\Q=\{\Q_{\i\btau}: \i \in \Omega, \btau \in W\}$.
For the corresponding ADMM algorithm, all variables have a closed-form update. However, the linear system corresponding to the $\X$-update is prohibitively expensive to invert.
More precisely, for the update $\X^{(t)} \to \X^{(t+1)}$, we need to solve a linear system of the form
\begin{equation}
\label{splitting_ls}
\Big(\B^\top\B + \I + \sum_{\i \in \Omega} \sum_{\btau \in W} \Pr_{\i\btau}^\top\Pr_{\i\btau}\Big) \X^{(t+1)} = \C,
\end{equation}
The coefficient matrix in \eqref{splitting_ls} is of size $n_h \times n_h$, where $n_h$ is the number of pixels. Hence, we cannot use a direct solver. 
Even for iterative solvers such as CG \cite{luenberger1984linear}, we have to apply the coefficient matrix (as an operator)  on an image of size $n_h \times L_s$ in each iteration. While the blur $\B$ and its transpose can be computed efficiently, the patch-extraction operation $\Pr_{\i\btau}^\top\Pr_{\i\btau}$ has to be performed for each pair of neighboring pixels---this has a high computational cost of $\mathcal{O}(n_h L_s |W|)$. Thus, the  straightforward splitting in \eqref{fusionoptim2} leads to an expensive $\X$ update, which slows down the overall ADMM algorithm. 
We now show how we get around this issue by expressing patch variations using convolutions and by using a novel form of variable splitting.

\begin{figure}
\centering
\includegraphics[width=0.95\linewidth]{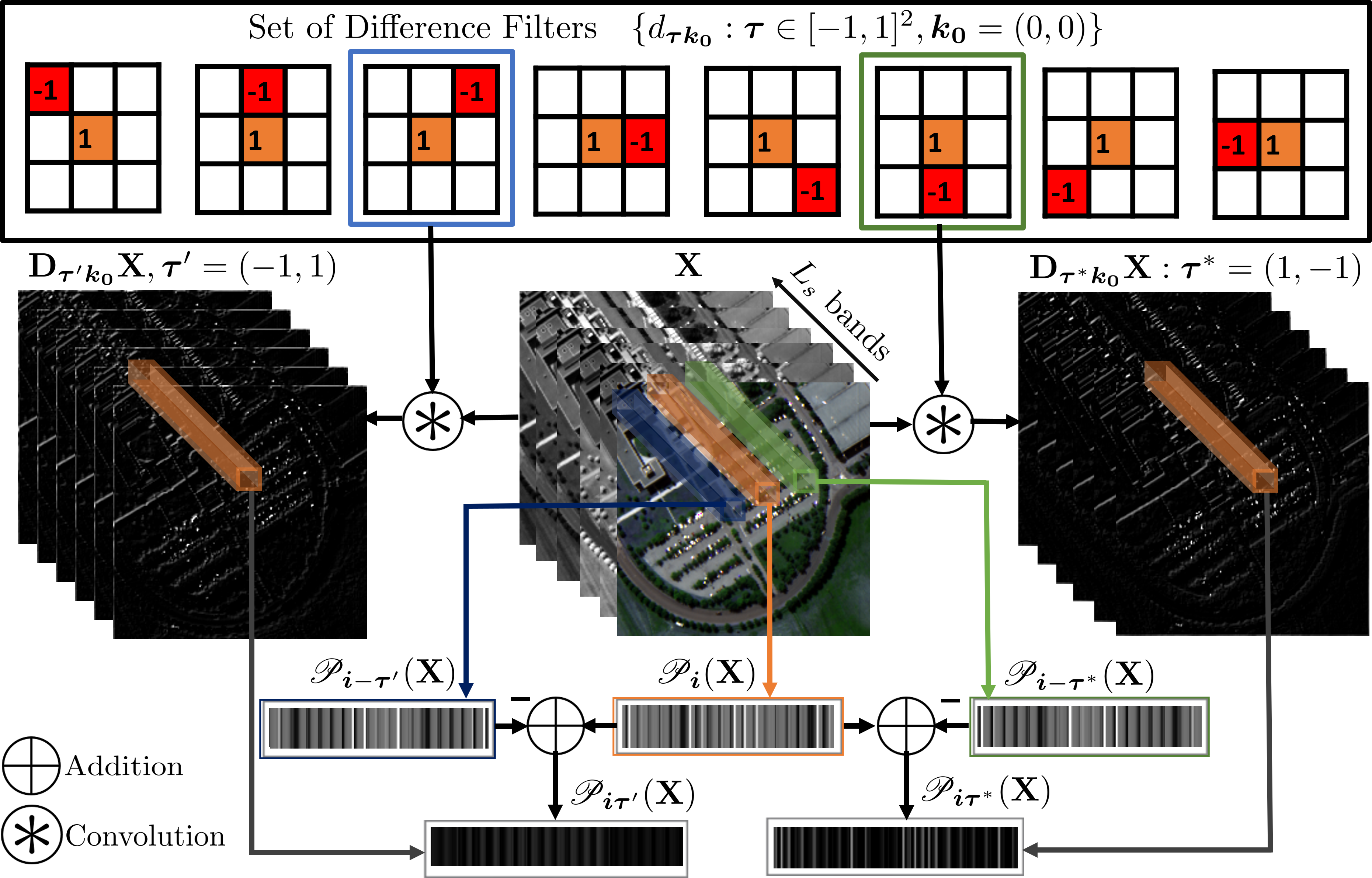}
\caption{Computation of patch variations $\Pr_{\i\btau'}(\X)$ and $\Pr_{\i\btau^*}(\X)$ using convolutions for a fixed pixel location $\i$ and two different shifts $\btau^\prime$ and $\btau^\ast$. 
}
\label{figillustr}
\end{figure}
To express the  patch variations in \eqref{reg1} using convolutions, for each $\btau \in W$ and $\k \in P$, we define the operator $\D_{\btau\k}: \Re^{n_h \times L_s}  \to  \Re^{n_h \times L_s}$ as
\begin{equation}
\label{differencemat}
\D_{\btau\k} \X(i,c) = X_c(\i - \k) - X_c(\i - \btau - \k), 
\end{equation}
where recall that $i \in [1,pq]$ is the row index of $\X$ corresponding to the pixel position $\i \in \Omega$. Also, recall that $X_c$ denotes the $c$-th band of $\X$. 

It is clear from definition \eqref{differencemat} that  $\D_{\btau\k}$ is a linear operator. In fact, $\D_{\btau\k}$ is a convolution operator. Indeed, if we define the filter $d_{\btau\k}: \Omega \to \Re$ to be 
$d_{\btau\k}(\i) = 1$ if $\i \equiv \k$, $d_{\btau\k}(\i) = -1$ if $\i \equiv \btau +\k$, and $d_{\btau\k}(\i) =0$ otherwise\footnote{We use $\i \equiv \k$ to mean $i_1 = k_1\! \!\mod p$ and $i_2 = k_2 \! \! \mod q$; similarly, $\i \equiv \btau +\k$ means $i_1= (\tau_1+k_1)   \! \!  \mod p$ and $i_2= (\tau_2+k_2)   \! \! \mod q$. Recall that $\Omega=[0,p-1]\times[0,q-1]$, the domain of $d_{\btau \k}$, is periodic.}, 
then we can write \eqref{differencemat} as
\begin{align*}
\D_{\btau\k} \X(i,c) = \sum_{\j} d_{\btau\k}(\j) X_c(\i-\j)=(d_{\btau\k} \circledast X_c)(\i),
\end{align*}
where $ \circledast $ denotes the two-dimensional circular convolution of $X_c$ with filter $d_{\btau\k}$ (recall that we impose periodic boundary conditions on $X_c$). We remark that the quantity on the left is the $(i,c)$-th entry of matrix $\D_{\btau\k} \X$, while the quantity on the right is output of the convolved image $d_{\btau\k} \circledast X_c$ at pixel position $\i$. Note that there are a total of $|W||P|$ filters, one for each $\btau \in W$ and $\k \in P$. Also note that, for a fixed $\btau \in W$, the filters are shifted versions of each other: if we let $\k_{\textbf{0}} = (0,0)$, then $(d_{\btau\k} \circledast \X_c)(\i) = (d_{\btau\k_{\textbf{0}}} \circledast \X_c)(\i - \k)$ for all $\k \in P \backslash \k_{\textbf{0}}$. This observation is used to simplify computations.

For fixed $\i$ and $\btau$, recall that  the patch variation $\Pr_{\i\btau}(\X)$ is the vector consisting of the samples $X_c(\i - \k) - X_c(\i -\btau -\k)$ for $(\k,c) \in \Lambda$. However, as per definition \eqref{differencemat}, this is just the vector representation of $\D_{\btau\k}\X(i,c)$ for $\k \in P$ and $c \in [1,L_s]$. These two different ways of computing patch variation is illustrated in figure \ref{figillustr}. 

In terms of the convolution filters, we can write 
\begin{equation*}
\lVert \Pr_{\i\btau}(\X) \rVert_1 =   \sum_{c=1}^{L_s}  \sum_{\k \in P}| \D_{\btau\k}\X(i,c)|,
\end{equation*}
and hence \eqref{reg1} as
\begin{align}
\label{regfilt1}
\phi(\X) 
=\sum_{\i \in \Omega}    \sum_{c=1}^{L_s} \sum_{\btau \in W}\sum_{\k \in P} \omega_{\i\btau} | \D_{\btau\k}\X(i,c) |.
\end{align}
We remark that $\i$ in the summation is the pixel position in the image domain $\Omega$ and the $i$ appearing in $\D_{\btau\k}\X(i,c)$  is the corresponding row index of the pixel in $\D_{\btau\k}\X$. 

\section{Optimization}
\label{solver}

The advantage with representation \eqref{regfilt1} will be clear in the context of the ADMM algorithm, where we will split the matrix variables $\{\D_{\btau\k}\X: \btau \in W, \k   \in P\}$. In particular, instead of the splitting in \eqref{fusionoptim2}, we consider the following equivalent formulation of \eqref{ouroptim}:
\begin{equation}
\begin{gathered}
\label{fusionoptim}
\min \ \frac{1}{2}\!\|\Y_\ell \!-\! \S\P_1\E\|^2 \!+\! \frac{\lambda_1}{2}\!{\|\Y_h \!- \!\P_2\E\R\|}^2 \!+\! \frac{\lambda_2}{2}\!g(\Q)  \\
\mbox{s.t.} \quad \P_1=\B\X, \quad \P_2=\X, \\ 
\D_{\btau\k}\X = \Q_{\btau\k} \quad (\btau \in W, \k   \in P),
\end{gathered}
\end{equation}
where $\X,\P_1,\P_2$ and $\Q=\{\Q_{\btau\k}: \btau \in W, \k \in P\}$ are the (primal) variables, and where, following \eqref{regfilt1}, we define
\begin{equation*}
g(\Q) = \sum_{\i \in \Omega} \sum_{c=1}^{L_s} \sum_{\btau \in W}   \sum_{\k \in P}  \  \omega_{\i\btau} |{\Q_{\btau\k} }(i,c) |. 
\end{equation*}
As explained next, the key advantage with formulation \eqref{fusionoptim} is that the subproblems in the corresponding ADMM algorithm have closed-form solutions.



Let $\BLambda_1$, $\BLambda_2$ and $\{\BSigma_{\btau\k}: \btau \in W, \k \in P\}$ be the dual variables for the equality constraints in \eqref{fusionoptim}. The augmented Lagrangian is given by
\begin{align*}
&\mathcal{L}_{\rho}(\X,\P_1,\P_2,\Q) =  \frac{1}{2}\|\Y_\ell - \S\P_1\E\|^2 + \frac{\lambda_1}{2}{\|\Y_h - \P_2\E\R\|}^2  \\ 
&+\frac{\lambda_2}{2} g\big( \Q) + \frac{\rho}{2}\|\P_1 - \B\X +\BLambda_1\|^2 + \frac{\rho}{2}\|\P_2 - \X +\BLambda_2\|^2 \\
  &+\frac{\rho}{2}\sum_{\btau\in W}\sum_{\k \in P}\|\Q_{\btau\k} - \D_{\btau\k}\X +\BSigma_{\btau\k}\|^2.
 \end{align*} 
where  $\rho>0$ is a penalty parameter \cite{boyd2011distributed}. 

Starting with initial variables $\P_1^{(0)}$, $\P_2^{(0)}$, $\Q_{\btau\k}^{(0)}$, $\BLambda_1^{(0)}$, $\BLambda_2^{(0)}$, $\BSigma_{\btau\k}^{(0)}$, we now work out the ADMM updates.

\subsection{$\X$ update}
The subproblem corresponding to $\X$ update is given by:
\begin{align}
\label{x_update_problem}
&\X^{(t+1)}=\underset{\X}{\mathrm{argmin}} \ \mathcal{L}_{\rho}(\X,\P_1^{(t)},\P_2^{(t)},\Q^{(t)})\nonumber  \\
&=\underset{\X}{\mathrm{argmin}} \ \frac{\rho}{2}\|\P_1^{(t)} - \B\X +\BLambda_1^{(t)}\|^2 +\frac{\rho}{2}\|\P_2^{(t)} - \X +\BLambda_2^{(t)}\|^2  \nonumber \\
&+ \frac{\rho}{2}\sum_{\btau\in W} \sum_{\k \in P} \|\Q_{\btau\k}^{(t)} - \D_{\btau\k}\X +\BSigma_{\btau\k}^{(t)}\|^2.
\end{align}
The objective in \eqref{x_update_problem} is strongly convex and differentiable. Setting its gradient to  zero, we obtain
\begin{equation}
\label{x_linear_system_main}
\Big(\I + \B^\top\B + \sum_{\btau \in W} \sum_{\k \in P} \D_{\btau\k}^\top\D_{\btau\k}\Big) \X^{(t+1)} =  \C^{(t)}, 
\end{equation}
where the matrix $\C^{(t)}$ is given by
\begin{equation*}
 \B^\top\! \big(\P_1^{(t)} + \boldsymbol{\Lambda}_1^{(t)}\big)+\P_2^{(t)}+\BLambda_2^{(t)} + \sum_{\btau \in W} \sum_{\k \in P} \D_{\btau\k}^\top\Big(\Q_{\btau\k}^{(t)}+\BSigma_{\btau\k}^{(t)}\Big).
\end{equation*}
The coefficient matrix in \eqref{x_linear_system_main} is positive definite and the linear system is solvable. We now show how this can be solved using the Discrete Fourier Transform (DFT). 

Let $X^{(t+1)}_c$ be the $c$-th band of $\X^{(t+1)}$ (i.e., $c$-th column of $\X^{(t+1)}$) and  $C_c^{(t)}$ be the $c$-th column of $\C^{(t)}$. Writing \eqref{x_linear_system_main} columnwise, we obtain for $c=1,\ldots,L_s:$
\begin{align}
\label{xupdateoper}
& \quad X^{(t+1)}_c + \check{b} \circledast b \circledast X^{(t+1)}_c \nonumber+ \\&  \sum_{\btau \in W} \sum_{\k \in P} \big(\check{d}_{\btau\k} \circledast d_{\btau\k} \circledast X^{(t+1)}_c \big) = C^{(t)}_c, 
\end{align}
where we recall that $b$ and $\check{b}$ are the filters associated with $\B$ and $\B^\top$, and  $d_{\btau\k}$ and $\check{d}_{\btau\k}$ are the filters associated with $\D_{\btau\k}$ and $\D_{\btau\k}^\top$.
The DFT of both sides of \eqref{xupdateoper} gives us
\begin{equation*}
\Big(1 + {|\widehat{b}(\bomega)|}^2 +\sum_{\btau \in W} \sum_{\k \in P} {|\widehat{d}_{\btau\k}(\bomega)|}^2\Big)  \widehat{X}_c^{(t+1)}(\bomega) = \widehat{C}^{(t)}_c(\bomega), 
\end{equation*}
where $ \widehat{X}_c^{(t+1)}$ and $ \widehat{C}^{(t)}_c$ are the DFTs of  $X_c^{(t+1)}$ and $C^{(t)}_c$. Moreover, since $|\widehat{d}_{\btau\k}(\bomega)| = |\widehat{d}_{\btau\k_{\boldsymbol{0}}}(\bomega)|$, we have
\begin{equation}
\label{inv}
\widehat{X}_c^{(t+1)}(\bomega) =\Big(1 + {|\widehat{b}(\bomega)|}^2 +|P|\sum_{\btau \in W} {|\widehat{d}_{\btau\k_{\boldsymbol{0}}}(\bomega)|}^2\Big)^{-1} \widehat{C}^{(t)}_c(\bomega).
\end{equation}
On taking the inverse DFT, we obtain $X_c^{(t+1)}$ for each $c$ and hence the desired update  $\X^{(t+1)}$. Note that the inverse in \eqref{inv} can be precomputed and stored. 

\subsection{$\P_1$ update}
The subproblem corresponding to $\P_1$ update is given by:
\begin{align}
\label{P1_update_problem}
&\P_1^{(t+1)}=\underset{\P_1}{\mathrm{argmin}} \ \mathcal{L}_{\rho}(\X^{(t+1)},\P_1,\P_2^{(t)},\Q^{(t)}) \nonumber \\
&=\underset{\P_1}{\mathrm{argmin}} \ \frac{1}{2}\|\Y_\ell - \S\P_1\E\|^2 + \frac{\rho}{2}\|\P_1 - \B\X^{(t+1)} +\BLambda_1^{(t)} \|^2.
\end{align}
The objective in \eqref{P1_update_problem} is strongly convex and differentiable in $\P_1$.  Setting its 
the gradient to zero, we obtain
\begin{equation}
\label{P1grad}
\S^\top \S \P_1^{(t+1)} \E\E^\top + \rho \P_1^{(t+1)} = \S^\top\Y_\ell\E^\top + \rho (\B\X^{(t+1)} - \BLambda_1^{(t)}) 
\end{equation}
Recall that $\S$ is the sampling matrix. It has the property that 
\begin{equation}
\label{maskprop}
\S^\top \S \ \S^\top = \S^\top \quad \mbox{and} \quad (\S^\top \S)^2 = \S^\top \S.
\end{equation}
Multiplying \eqref{P1grad} by $\S^\top \S$ and $(\I-\S^\top \S)$ and using \eqref{maskprop}, we obtain
\begin{equation*}
\S^\top \S \ \P_1^{(t+1)} (\E\E^\top+\rho \I) = \S^\top \S \ (\rho\B\X^{(t+1)}+\S^\top\Y_\ell\E^\top-\rho\BLambda_1^{(t)}),
\end{equation*}
and
\begin{equation*}
(\I - \S^\top \S) \ \P_1^{(t+1)} = (\I - \S^\top \S) \ (\B\X^{(t+1)}-\BLambda_1^{(t)}).
\end{equation*}
Therefore, we can write
\begin{align}
\label{Qh_update}
&\P_1^{(t+1)} = \S^\top \S \ \P_1^{(t+1)} + (\I - \S^\top \S) \ \P_1^{(t+1)} \nonumber \\
&=\S^\top \S \ (\rho\B\X^{(t+1)}+\S^\top\Y_\ell\E^\top-\rho\BLambda_1^{(t)})\nonumber (\E\E^\top+\rho \I)^{-1}\\&  + (\I - \S^\top \S) \ (\B\X^{(t+1)}-\BLambda_1^{(t)}). 
\end{align}
Note that $\S^\top \S$ is diagonal with $1$ in positions corresponding to the sampled pixels and $0$ elsewhere \cite{simoes2015convex} (recall that $\S$ is the spatial sampling operator). Since the matrix $\E\E^\top + \rho\I$ is of size $L_s \times L_s$, its inverse can be precomputed and stored.
\subsection{$\P_2$ update}
The subproblem corresponding to $\P_2$ update is given by:

\begin{align}
\label{P2_update_problem}
&\P_2^{(t+1)}=\underset{\P_2}{\mathrm{argmin}} \ \mathcal{L}_{\rho}(\X^{(t+1)},\P_1^{(t+1)}\P_2,\Q^{(t)}) \nonumber  \\&\hspace{-0.25cm}=\underset{\P_2}{\mathrm{argmin}} \ \frac{\lambda_1}{2}{\|\Y_h - \P_2\E\R\|}^2 +\frac{\rho}{2}\|\P_2 + \BLambda_2^{(t)}- \X^{(t+1)} \|^2. 
\end{align}
Setting $\nabla_{\P_2} \mathcal{L}_{\rho}(\X^{(t+1)},\P_1,\P_2^{(t)},\Q^{(t)}) = \mathbf{0}$, we obtain the  closed-form solution:
\begin{equation}
\label{Qm_update}
\P_2^{(t+1)} = (-\BLambda_2^{(t)}+\X^{(t+1)}+\lambda_1/\rho \ \Y_h\R^\top\E^\top)\F,
\end{equation}
where $\F = (\I+\lambda_1/\rho \ \E\R\R^\top\E^\top)^{-1}$ is small ($L_s \times L_s$) and hence can be precomputed and stored.

\subsection{$\Q$ update}
The subproblem corresponding to the $\Q$ update is given by:

\begin{equation*}
\Q^{(t+1)}=\underset{\Q}{\mathrm{argmin}} \ \mathcal{L}_{\rho}(\X^{(t+1)},\P_1,\P_2^{(t+1)},\Q)
\end{equation*}


Using the separable structure, we can break this up into smaller subproblems. Namely, if we let
$$\tilde{\Q}_{\btau\k}^{(t)} = \D_{\btau\k}\X^{(t+1)}-\BSigma_j^{(t)},$$
then, for $\btau \in W, \k \in P$, $\forall i\in [1,n_h], \ c \in [1,L_s],$
\begin{align}
\label{Qj_update_problem}
\Q_{\btau\k}^{(t+1)}(i,c)= &\underset{\Q_{\btau\k}}{\mathrm{argmin}} \ \Big\{ \big| {\Q_{\btau\k}}(i,c) \big| \nonumber \\&+  \frac{\rho}{2\lambda_2 \ \omega_{\i\btau}}    \Big(\Q_{\btau\k}(i,c) - \tilde{\Q}_{\btau\k}^{(t)}(i,c) \Big)^2 \Big\}.
\end{align}

For each $\btau \in W$ and $\k \in P$, the matrix $\Q_{\btau\k}$ is updated using soft-thresholding operation (on account of the $\ell_1$-norm in the regularizer). The exact update is as follows:
\begin{equation}
\label{Qj_main}
{\Q_{\btau\k} }^{(t+1)} = \mathcal{S} \Big(\D_{\btau\k}\X^{(t+1)}-\BSigma_{\btau\k}^{(t)}, \lambda_2  \omega_{\i\btau} / \rho \Big),
\end{equation}
where $\mathcal{S}(x , \mu)=\mbox{sign}(x)\max(|x|-\mu,0)$ is the soft thresholding of $x$ with threshold $\tau$; note that $\mathcal{S}$ is applied componentwise on the matrix. 

\subsection{Dual updates}

The dual variables are updated as follows:
\begin{align}
&\BLambda_1^{(t+1)} = \BLambda_1^{(t)} - (\B\X^{(t+1)} - \P_1^{(t+1)}), \label{dualvar1_main} \\
&\BLambda_2^{(t+1)} = \BLambda_2^{(t)} - (\X^{(t+1)} - \P_2^{(t+1)}), \label{dualvar2_main} \\
&\BSigma_{\btau\k}^{(t+1)} = \BSigma_{\btau\k}^{(t)} - (\D_{\btau\k}\X^{(t+1)} - \Q_{\btau\k}^{(t+1)}). \label{dualvar3_main}
\end{align}

Recall that \eqref{fusionoptim} is a convex program. As is well-known, under appropriate technical conditions, the iterates generated by the ADMM algorithm are guaranteed to converge to the global optimum \cite{eckstein1992douglas}. For our algorithm, we have the following convergence result (see Appendix for proof).
\begin{theorem}
\label{maintheorem}
Assume that the original problem \eqref{ouroptim} is solvable, i.e., there exists $\X^*$ where the minimum is attained. Then, for any arbitrary initialization and any $\rho>0$, the sequence of updates  $(\X^{(t)})_{t \geqslant 0}$ generated using \eqref{x_linear_system_main}, \eqref{Qh_update}, \eqref{Qm_update}, \eqref{Qj_main},  \eqref{dualvar1_main}, \eqref{dualvar2_main} and \eqref{dualvar3_main} converges to $\X^*$.
\end{theorem}

\begin{figure*}
\centering
\includegraphics[width=0.7\linewidth]{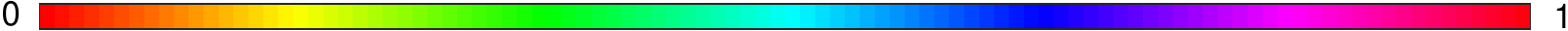}
\\
\subfloat[Ground truth.]{\includegraphics[width=0.136\linewidth]{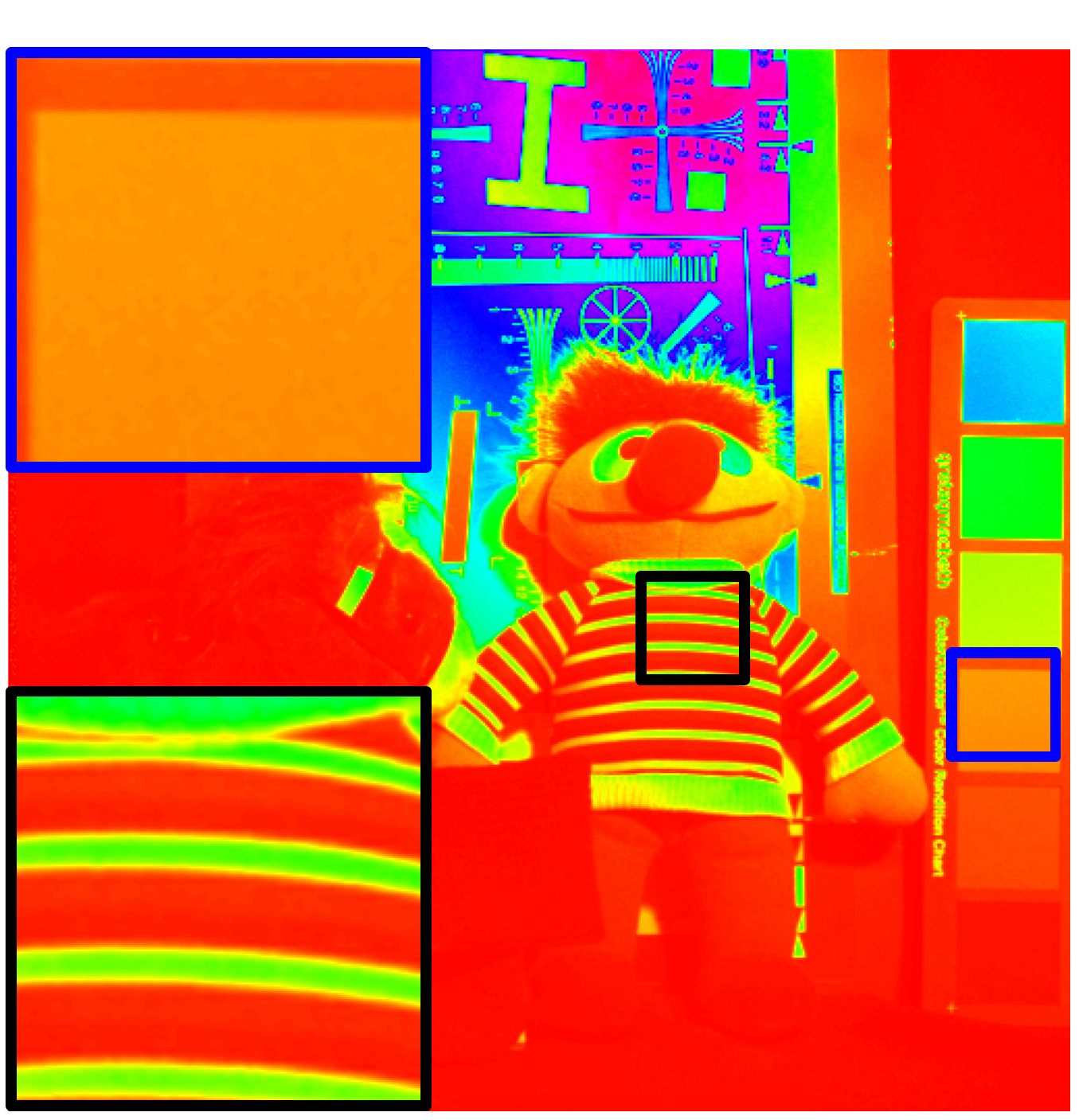}} \hspace{0.05mm}
\subfloat[Bicubic.]{\includegraphics[width=0.136\linewidth]{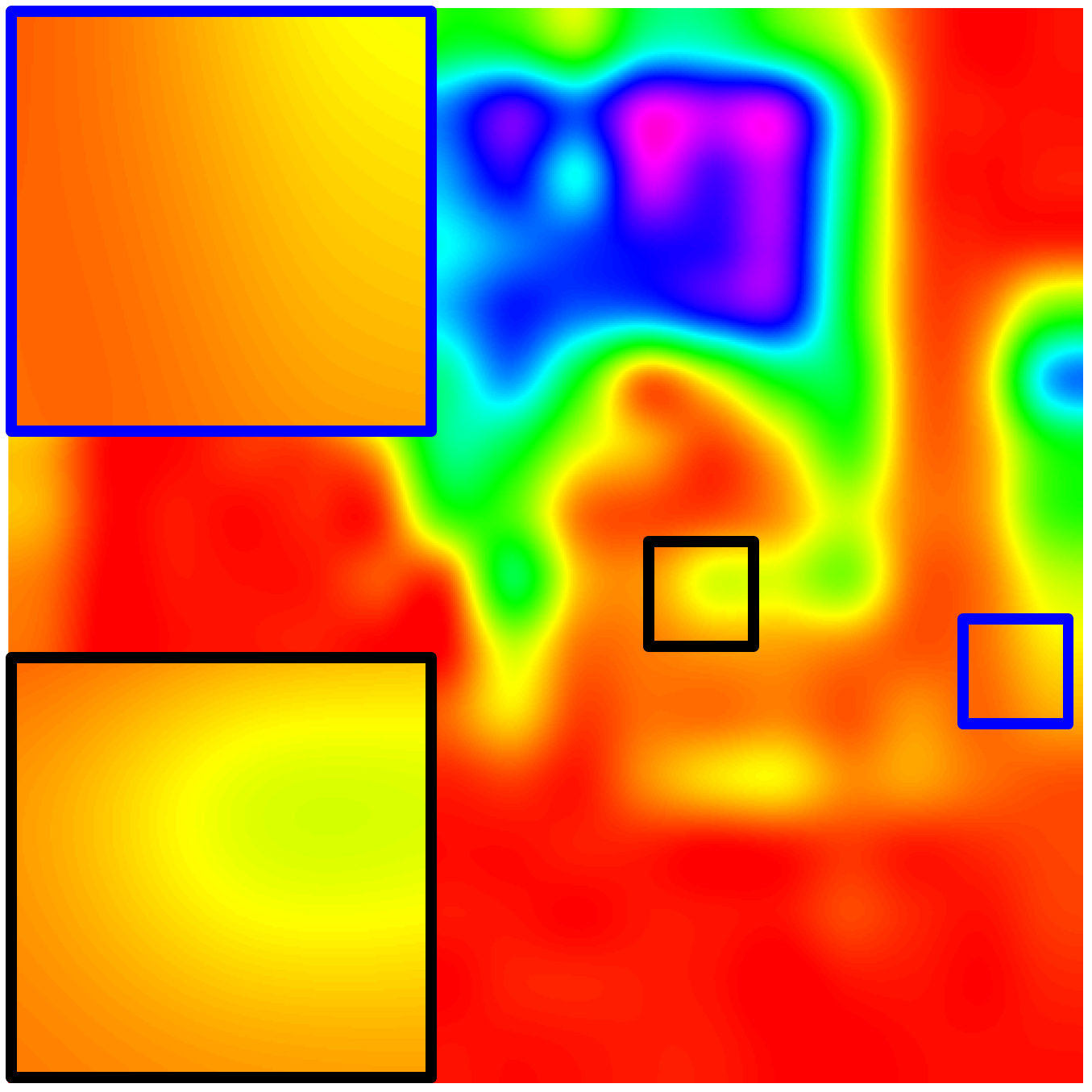}} \hspace{0.05mm}
\subfloat[GSA.]{\includegraphics[width=0.136\linewidth]{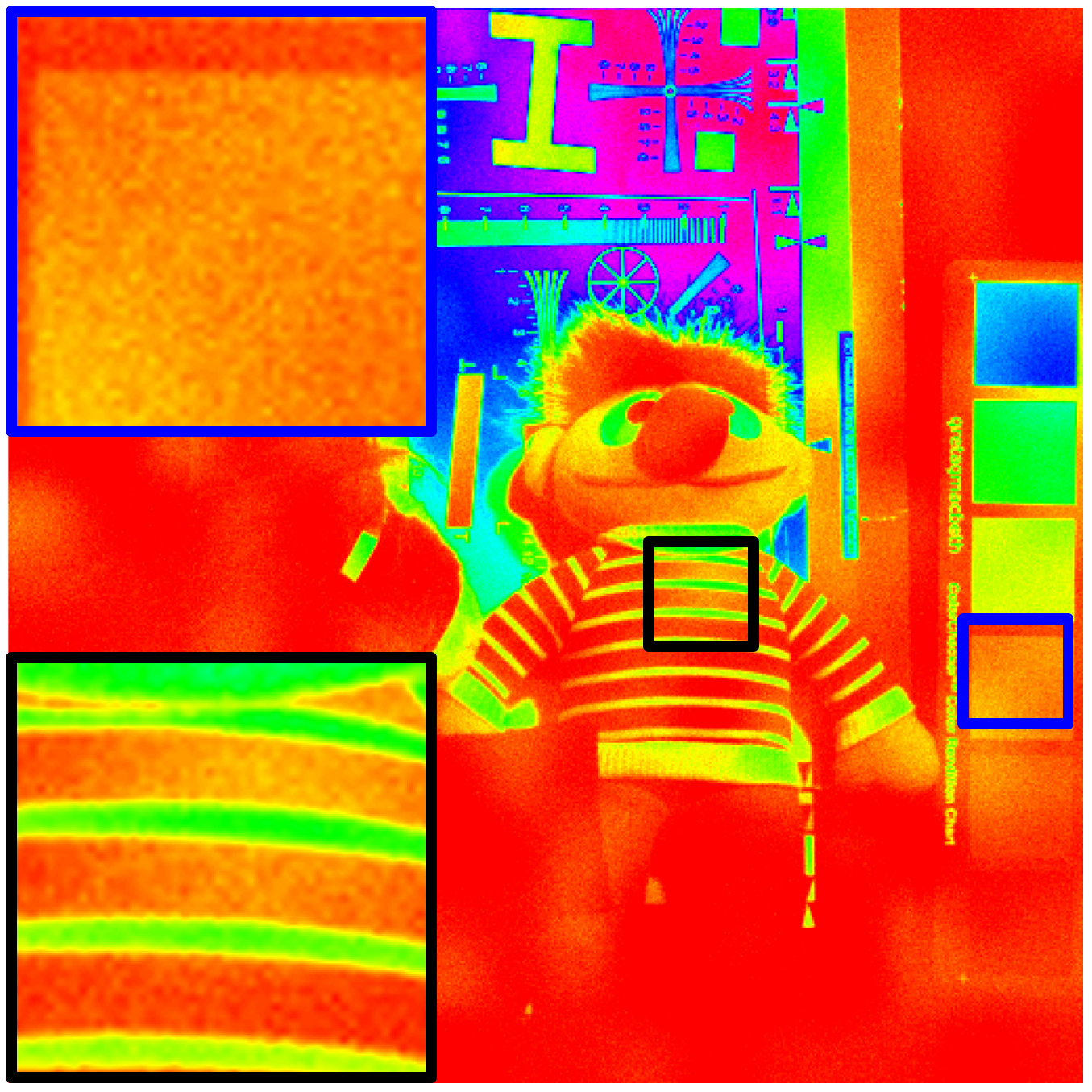}} \hspace{0.05mm}
\subfloat[CNMF.]{\includegraphics[width=0.136\linewidth]{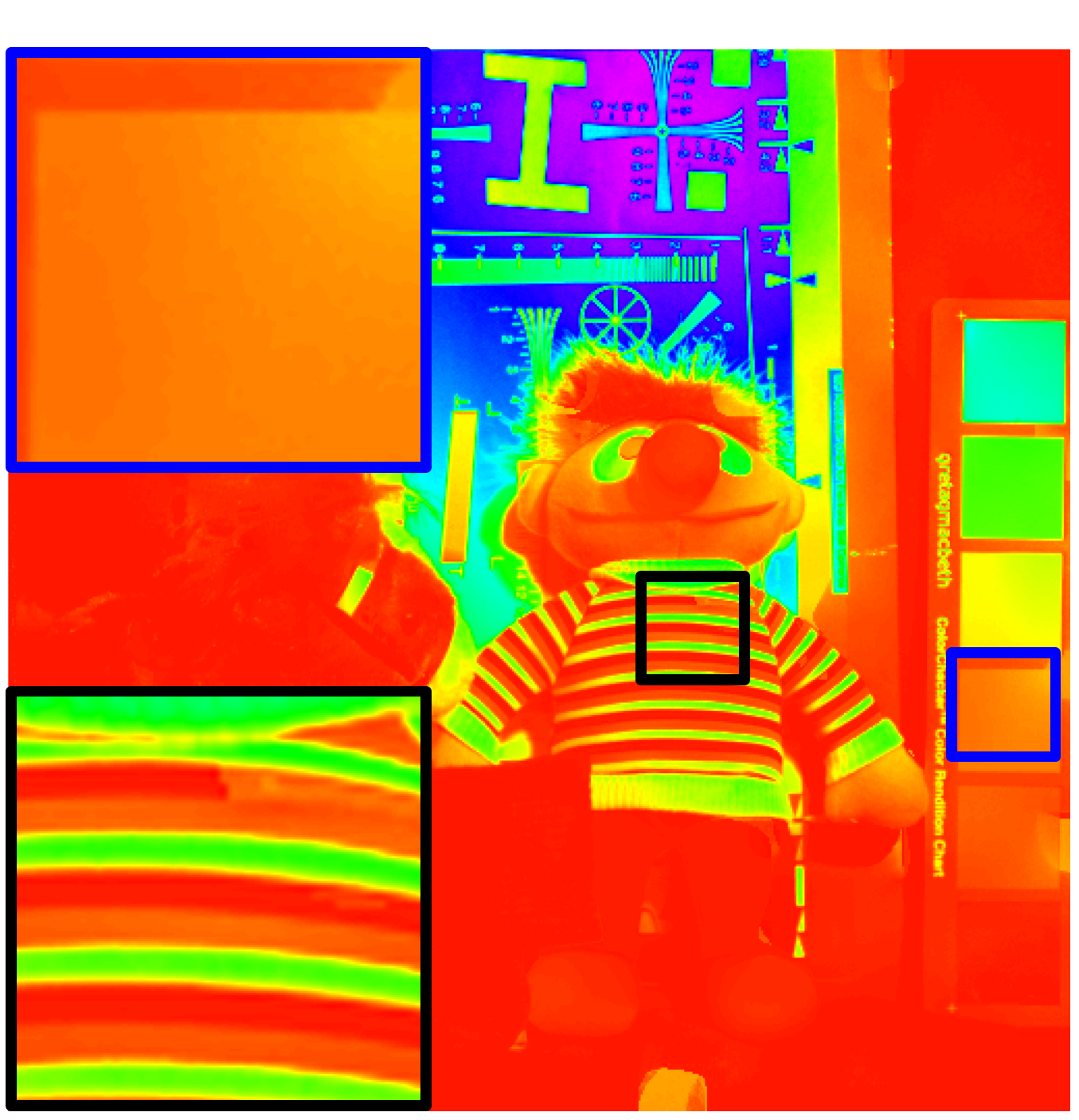}} \hspace{0.05mm}
\subfloat[JSU.]{\includegraphics[width=0.136\linewidth]{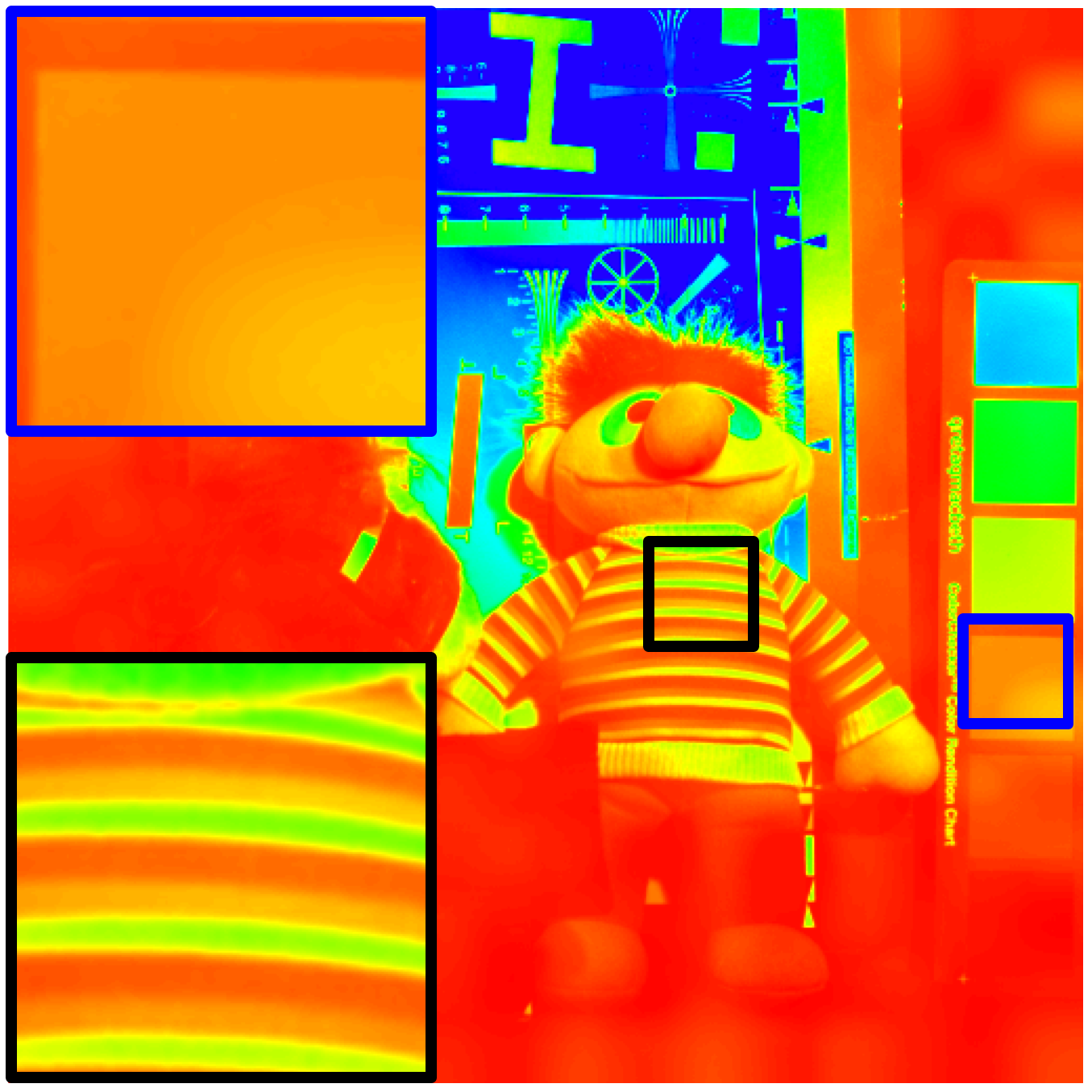}}  \hspace{0.05mm}
\subfloat[FUSE.]{\includegraphics[width=0.136\linewidth]{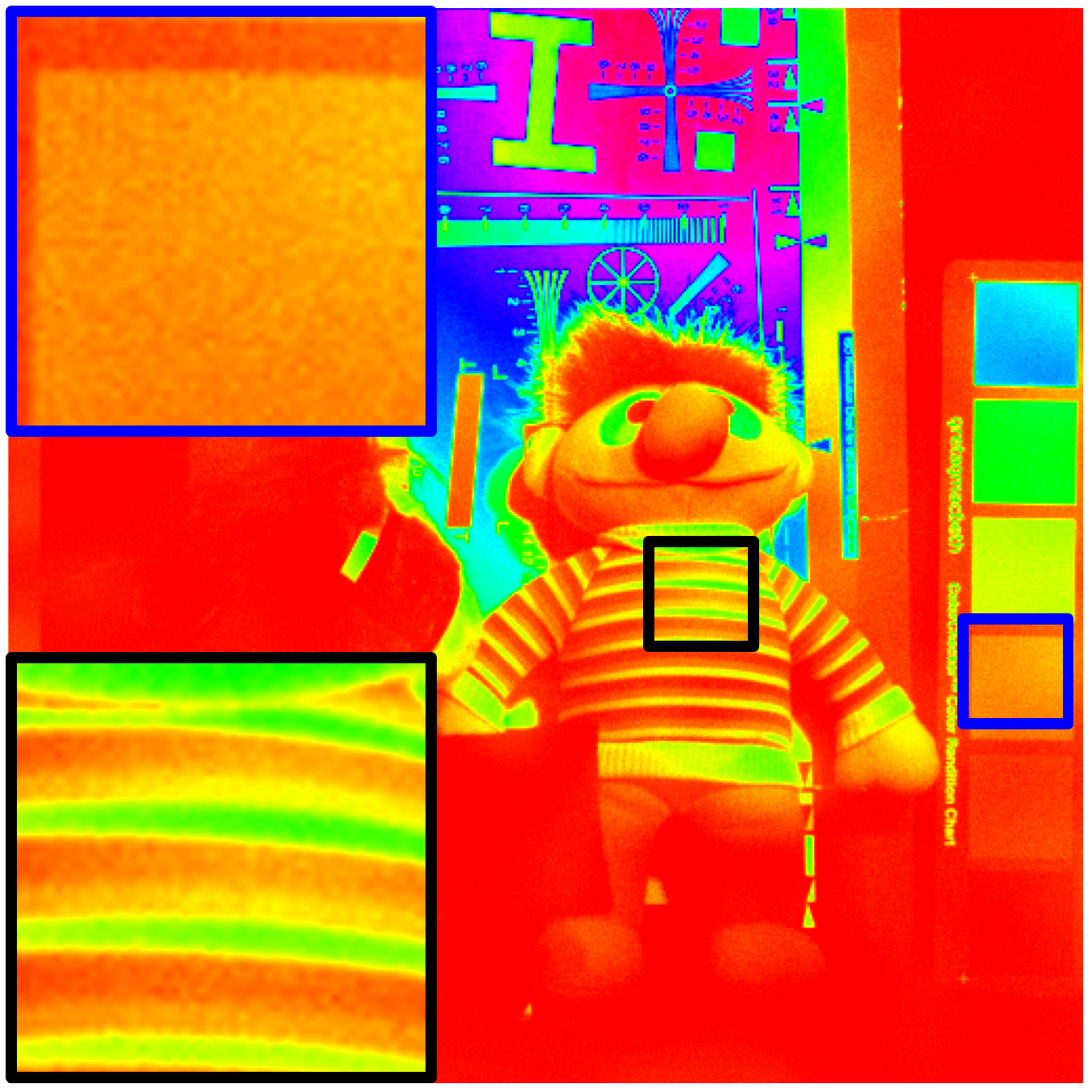}}  \hspace{0.05mm}
\subfloat[Sparse.]{\includegraphics[width=0.136\linewidth]{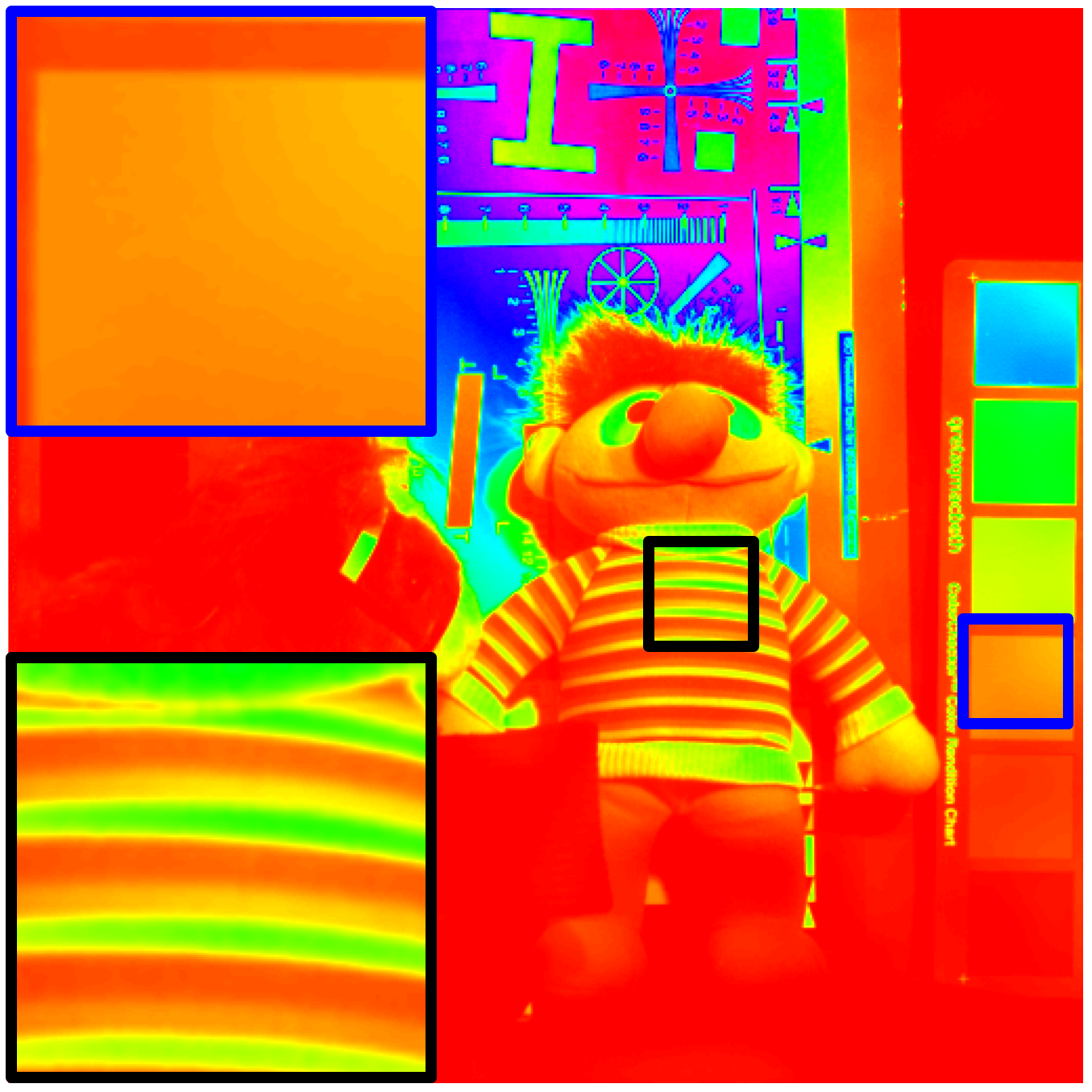}}  \hspace{0.05mm}\\
\subfloat[GLPHS.]{\includegraphics[width=0.136\linewidth]{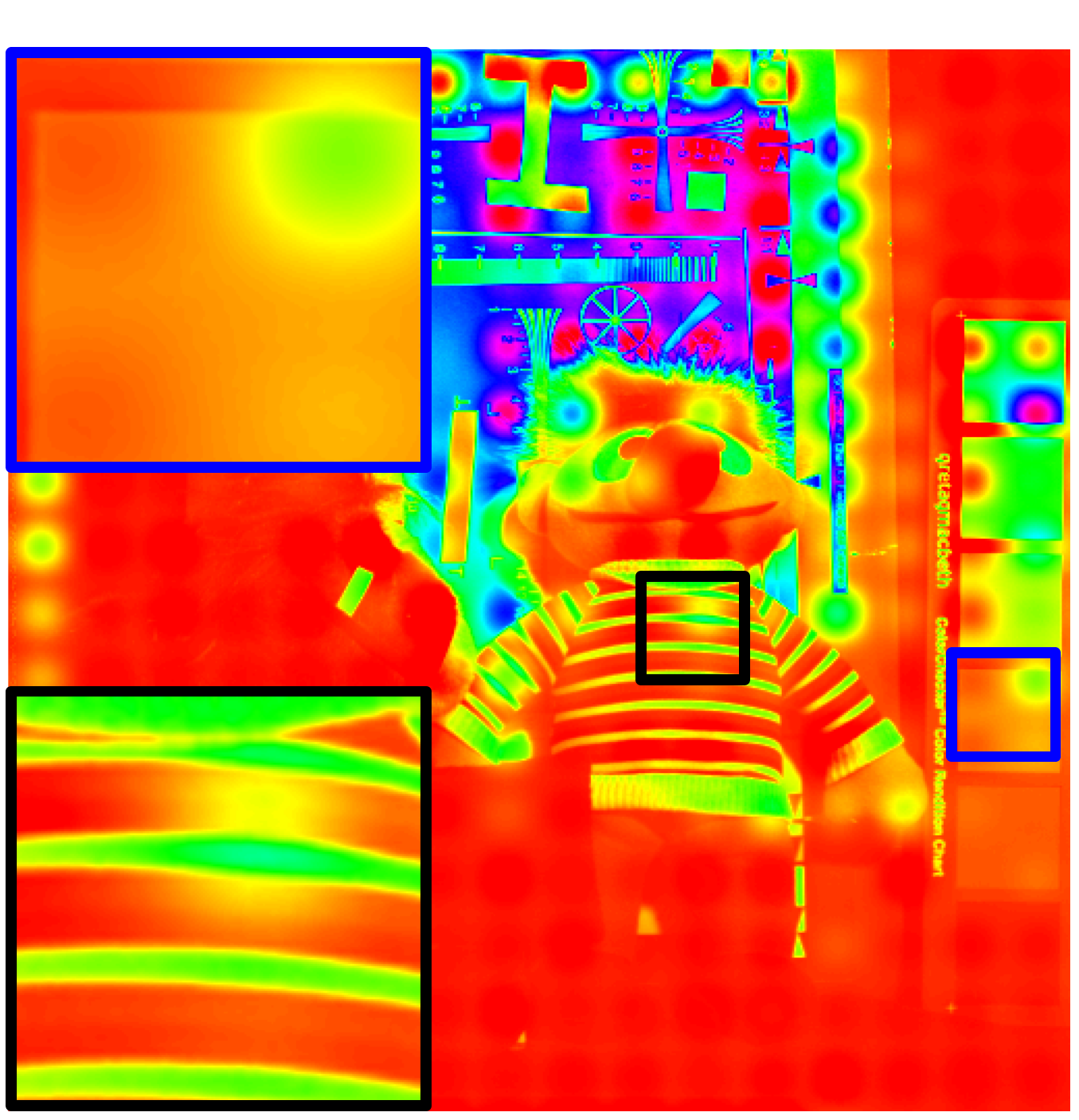}} \hspace{0.05mm}
\subfloat[MSMM.]{\includegraphics[width=0.136\linewidth]{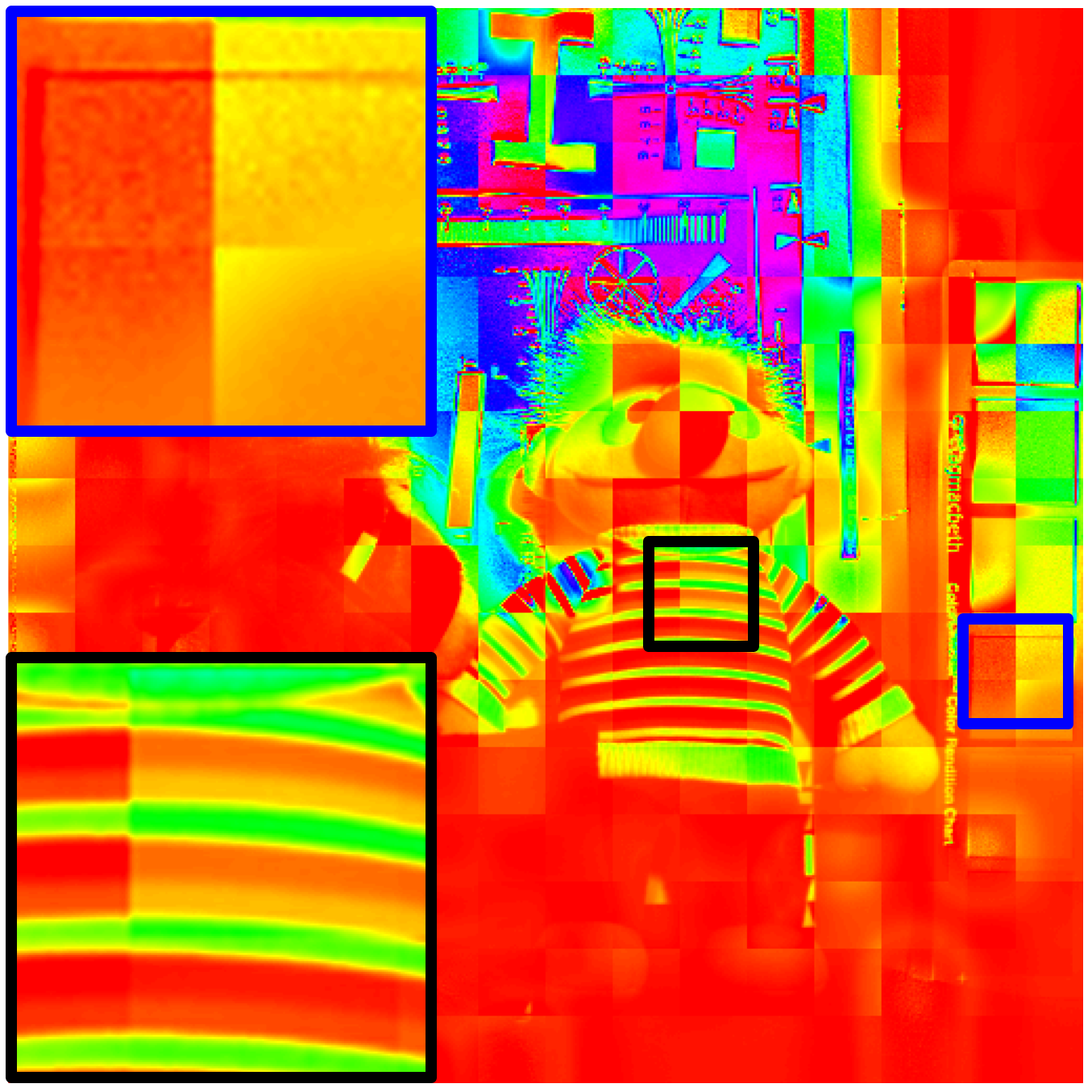}} \hspace{0.05mm}
\subfloat[NLSTF.]{\includegraphics[width=0.136\linewidth]{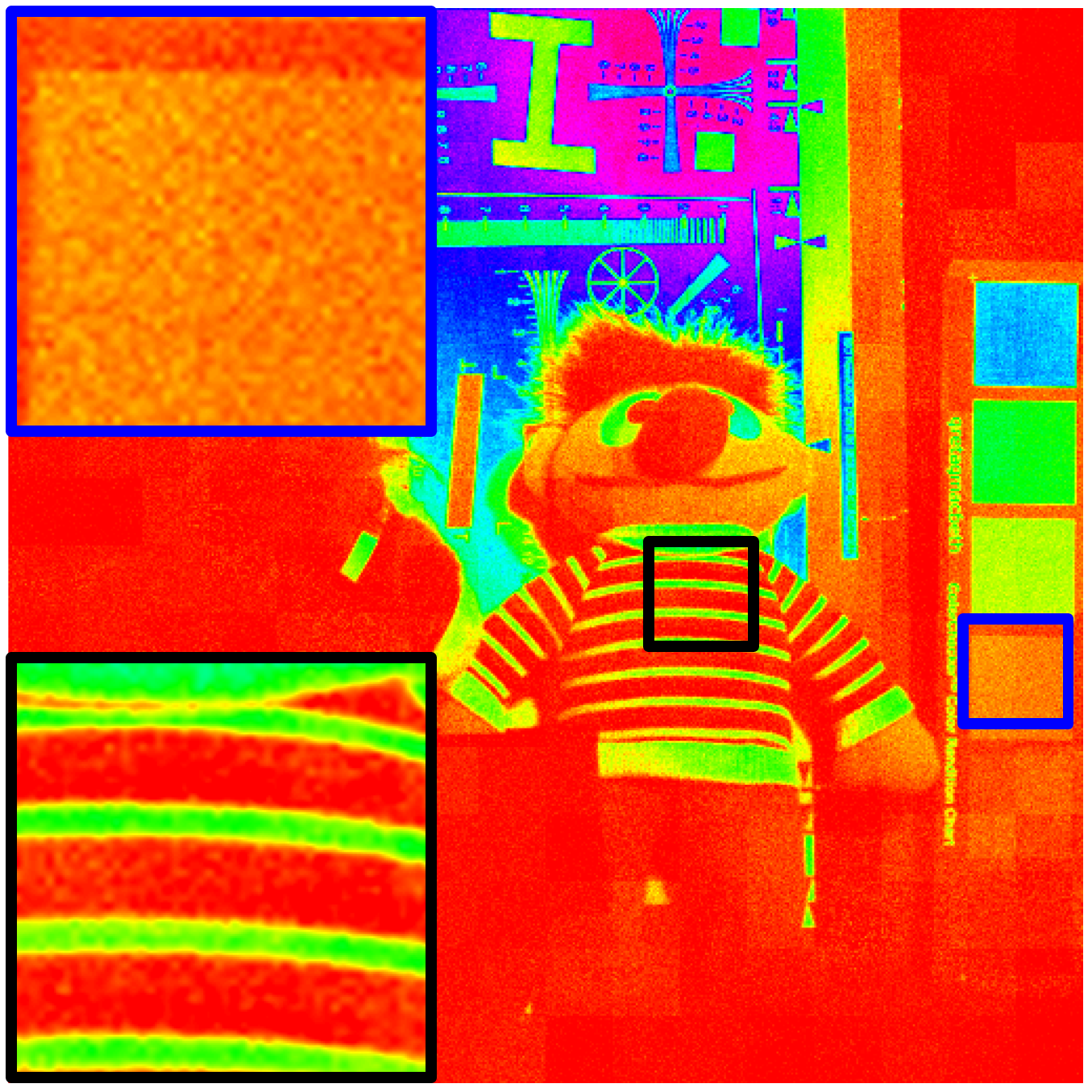}} \hspace{0.05mm}
\subfloat[SMBF.]{\includegraphics[width=0.136\linewidth]{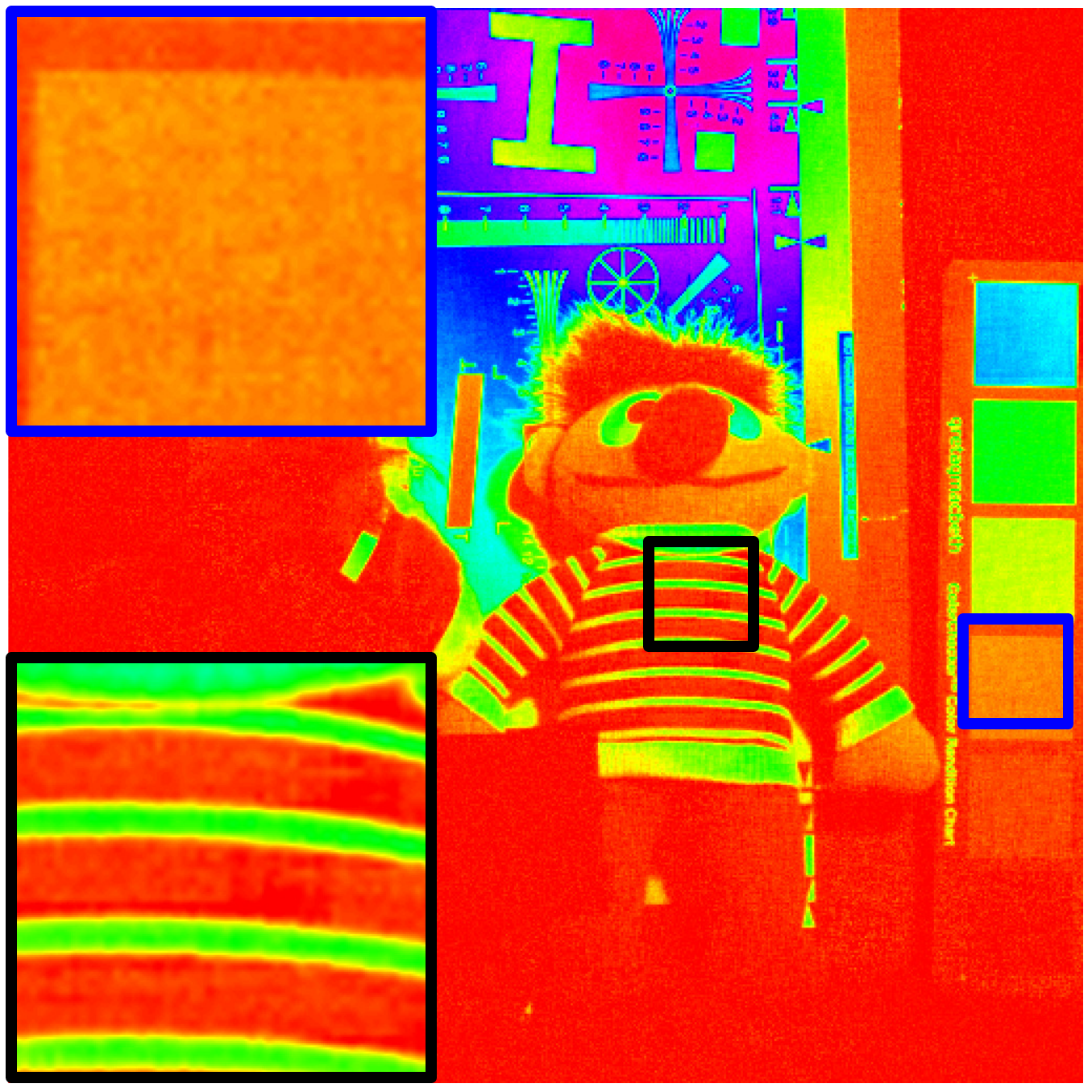}}  \hspace{0.05mm}
\subfloat[HySure.]{\includegraphics[width=0.136\linewidth]{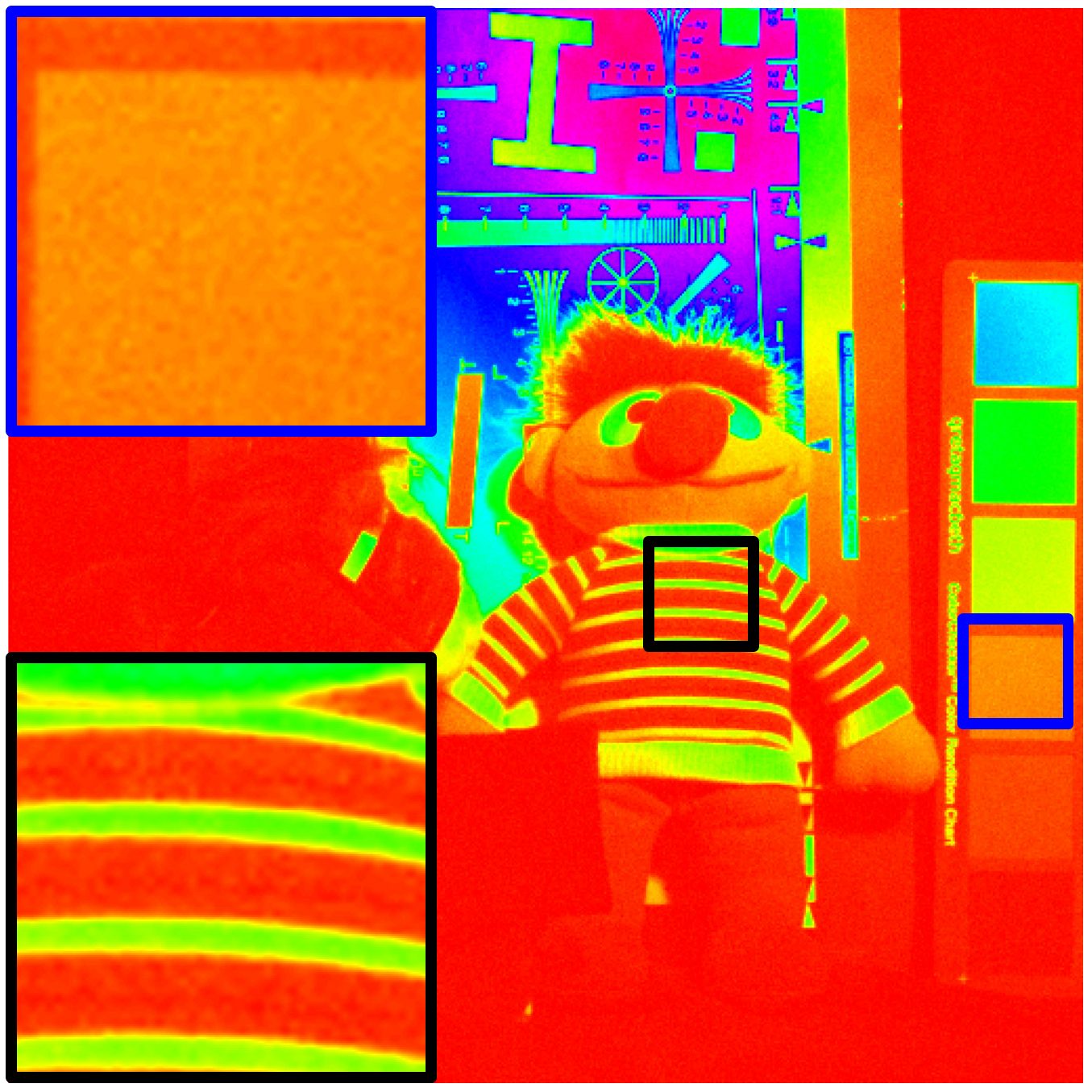}}  \hspace{0.05mm}
\subfloat[MHF.]{\includegraphics[width=0.136\linewidth]{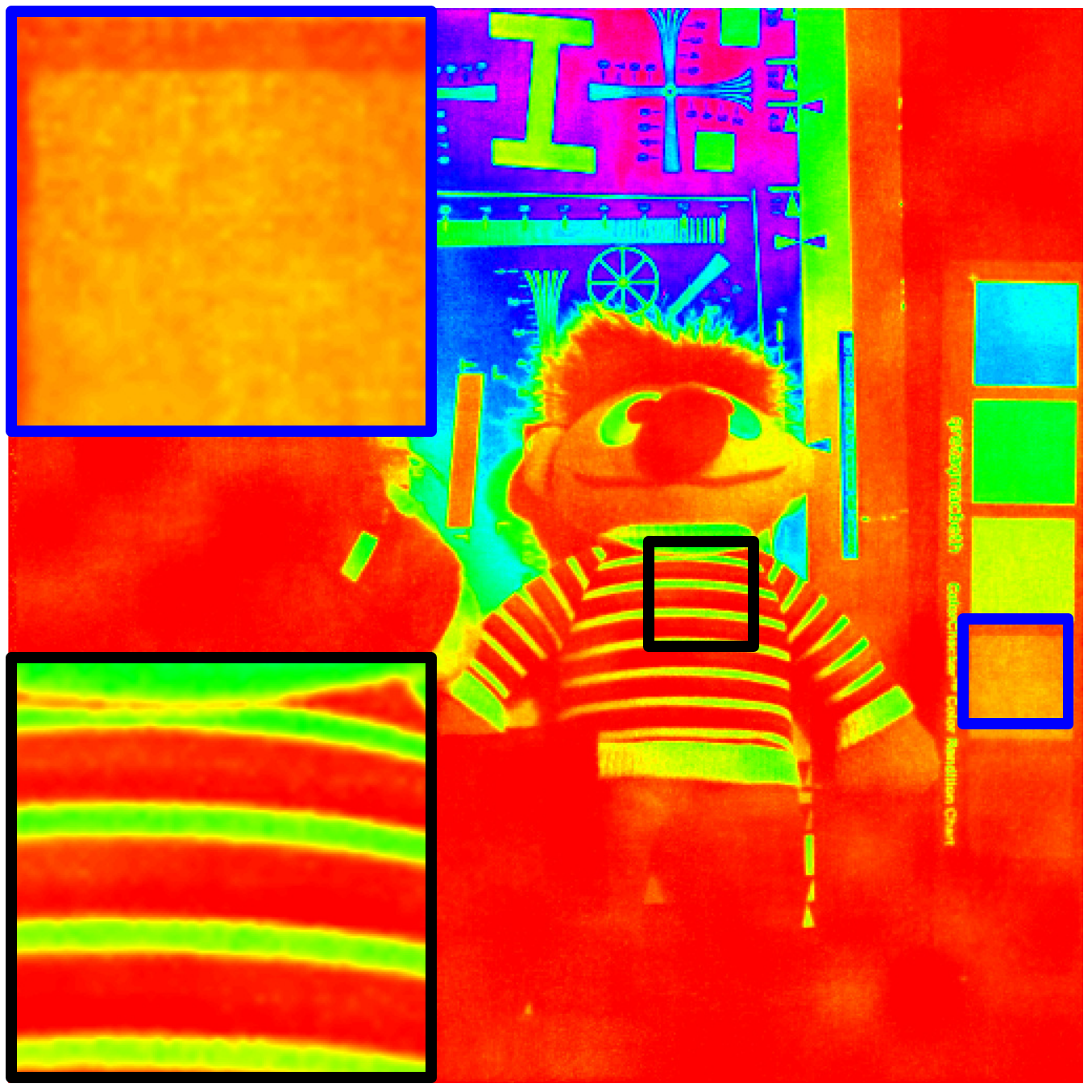}}  \hspace{0.05mm}
\subfloat[NLPR.]{\includegraphics[width=0.136\linewidth]{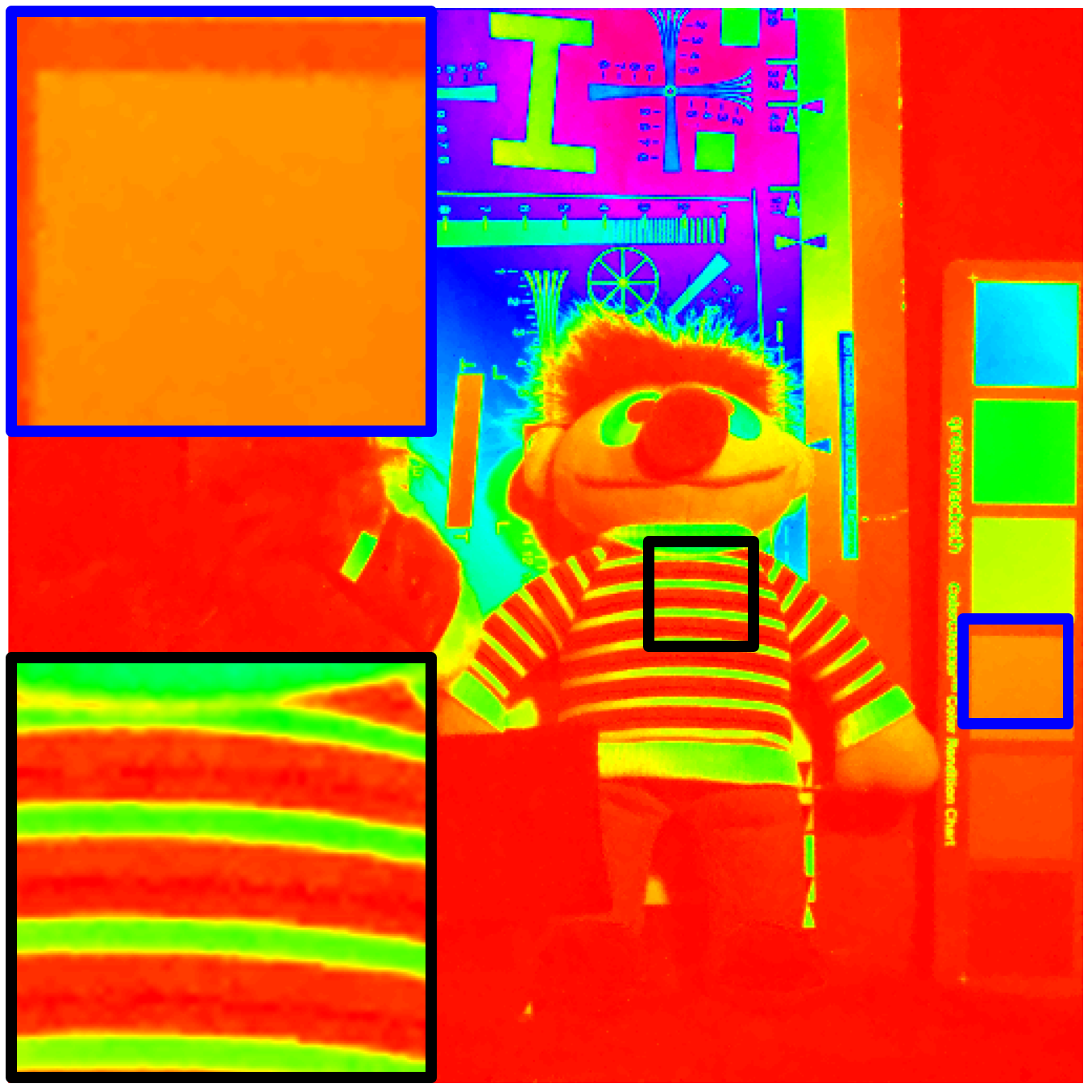}}  \hspace{0.05mm}
\caption{Fusion results for $\textbf{chart and stuffed toy}$ from the CAVE dataset at $\text{SNR}_h$ and $\text{SNR}_\ell$ of $30$ dB (this is the $16$th band). The proposed method (NLPR) is able to superresolve the multiband image while simultaneously preserving the spectral and spatial quality.}
\label{cave_results}
\end{figure*}

\begin{table*}
\centering
\scalebox{0.98}{
\begin{tabularx}{\textwidth}{c |*{13}{Y}}
\hline

Methods     & GSA    & CNMF & JSU   & FUSE  & Sparse & GLPHS  & MSMM   & HySure  & NLSTF  &SMBF  & MHF & NLPR\\ 

\hline

\hline
RMSE  &   $0.0272$ &   $0.0316$  &  $0.0426$ &   $0.0268$   & $0.0270$   & $0.0683$  &  $0.0765$  & $0.0165$  &  $0.0163$ &   $0.0155$  &  $0.0194$  &  $\textbf{0.0129}$ \\

ERGAS  &  $0.6506$   & $0.7819$  &  $1.0521$ &   $0.6383$  &  $0.6435$  &  $1.5938$  &  $1.7649$ &   $0.4013$  &  $0.3793$ &   $0.3751$   & $0.4632$   & $\textbf{0.3088}$ \\

SAM &  $24.685$ &  $15.729$  & $11.975$  & $19.401$ &  $18.605$  & $14.060$  & $16.391$  & $14.265$ &  $16.647$  & $17.043$ &  $13.993$ &   $\textbf{9.9723}$ \\

UIQI &   $0.6315$  &  $0.6619$   & $0.6581$ &   $0.6882$  &  $0.7275$  &  $0.5625$  &  $0.5488$  &  $0.8119$  &  $0.6572$  &  $0.7257$  &  $0.6473$  &  $\textbf{0.8192}$ \\

PSNR  &  $31.773$ &  $30.455$ &  $28.394$  & $31.860$  & $31.919$  & $24.466$  & $24.207$  & $35.838$  & $35.680$  & $36.341$ &  $34.365$ &  $\textbf{37.729}$\\
SSIM  &  $0.7608$ &  $0.8797$ &  $0.8555$  & $0.8225$  & $0.8810$  & $0.8326$  & $0.7795$  & $0.9400$  & $0.8578$  & $0.9167$ &  $0.8407$ &  $\textbf{0.9709}$
\\ \hline
\hline
\end{tabularx}}
\caption{Performance comparison for the CAVE dataset (averaged over $5$ images) using standard quality metrics.}
\label{objectivemeasure_cave}
\end{table*}

\subsection{Computational details}
\label{convergence}

Unlike the $\X$ update for \eqref{splitting_ls}, which is performed iteratively and is expensive,  the $\X$ update in \eqref{x_linear_system_main} can be computed efficiently.  Indeed, since $\B$ and $\D_{\btau\k}$ (and their transposes) are circular convolutions, they can be jointly diagonalized in a Fourier basis. Let $\widehat{b}$ and $ \widehat{d}_{\btau\k}$ denote the discrete Fourier transform (DFT) of filters $b$ and $d_{\btau\k}$. Recall that the DFT of the $\X$ update is given by
\begin{equation*}
\widehat{X}_c^{(t+1)}(\bomega) =\!\Big(1 + {|\widehat{b}(\bomega)|}^2\! + \!\sum_{\btau \in W} \sum_{\k \in P} |\widehat{d}_{\btau\k}(\bomega)|^2\Big)^{-1} \!\widehat{C}^{(t)}_c(\bomega),
\end{equation*}
where $C^{(t)}_c$ is the image representation of $\C^{(t)}(:,c)$ in \eqref{x_linear_system_main} and $ \widehat{C}^{(t)}_c$ is its DFT. 

Thus, the $\X$ update can be performed using two FFTs that has complexity $\mathcal{O}(|W ||P|L_s n_h \log(n_h))$. Note that the first term on the right involving the DFT of the filters can be precomputed. 
Also, the matrix inverses in \eqref{Qh_update} and \eqref{Qm_update} can be precomputed  and stored, since the matrix size is $L_s \times L_s$ where $L_s$ is in the range $10\mbox{-}20$ for HS-MS fusion and hyperspectral sharpening, and in the range $4\mbox{-}8$ for pansharpening. For updates \eqref{Qh_update} and \eqref{Qm_update}, the complexity is $\mathcal{O}(L_s n_h \log(n_h))$ for constructing the matrices and $\mathcal{O}(L_s^3)$ for inverting the matrix. The complexity of the thresholding operations in \eqref{Qj_main} is $\mathcal{O}(|W| |P| L_s n_h )$. Clearly, the cost per iteration of the proposed algorithm is dominated by the $\X$ update which we perform efficiently using FFTs. For example, on multiband image of size $200 \times 200 \times 20$, solving \eqref{splitting_ls}  using CG takes $127$s ($50$ CG iterations), whereas \eqref{x_linear_system_main}  takes just $135$ms ($940\times$ speedup).

\section{Experiments}
\label{experiments}

We test our algorithm for HS+MS fusion and pansharpening (PAN+MS fusion) using twelve pairs of $\Y_\ell\mbox{-}\Y_h$ images.
We conducted experiments on three diverse multiband datasets at different SNR ($\text{SNR}_h$ and $\text{SNR}_\ell$) levels \cite{simoes2015convex}. 
We compare with state-of-the-art fusion techniques including a recent deep learning method \cite{xie2019multispectral}.
We report some representative results and discuss them; additional results can be found in the supplement. We also perform an ablation study of our method and  compare it with  existing nonlocal regularizers.

\begin{table*}
\centering
\begin{adjustbox}{max width=30cm}
\scalebox{0.975}{
\begin{tabularx}{\textwidth}{c | c |*{12}{Y}}
\hline

Methods  & Dataset   & GSA    & CNMF & JSU   & FUSE  & Sparse & GLPHS  & MSMM   & HySure  & NLSTF  &SMBF  & NLPR\\
\hline
\hline
RMSE  & & $0.0187$ & $0.0255$ & $0.0185$ & $0.0186$ & $0.0177$  & $0.0310$  & $0.0318$ & $0.0152$ & $0.0205$  & $0.0174$ & $\mathbf{0.0125}$ \\

ERGAS & & $2.9978$ & $4.1269$ & $2.6849$ & $2.6781$ & $2.5651$ & $5.1250$ & $5.2430$ & $2.5725$ & $3.4042$ & $2.8627$  & $\mathbf{1.8432}$  \\    

SAM & Pavia & $5.1151$ & $5.2241$ & $3.7204$ & $4.0816$ & $3.2950$ & $6.2572$ & $5.8747$ & $3.4064$ & $6.7796$ & $5.4051$ & $\mathbf{2.8760}$ \\

UIQI  & & $0.9722$ & $0.9576$ & $0.9773$ & $0.9791$ & $0.9817$ & $0.9370$ & $0.9317$ & $0.9820$ & $0.9560$ & $0.9677$  &$\mathbf{0.9879}$ \\      

PSNR  &  & $34.583$ & $31.872$ & $34.669$ & $34.613$ & $35.053$ & $30.169$ & $29.955$  & $36.234$  & $33.747$ & $35.193$ &  $\mathbf{38.039}$  
\\ 
SSIM&& $0.7420$ & $0.7604$ & $0.9146$ & $0.8628$ & $0.8907$  & $0.7760$  & $0.6777$ & $0.9448$ & $0.8567$  & $0.9206$ & $\mathbf{0.9710}$ \\
\hline
\hline
RMSE  & & $0.0416$ & $0.0423$ & $0.0252$ & $0.0361$ & $0.0337$ & $0.0437$ & $0.0543$ & $0.0181$ & $0.0255$ & $0.0198$ & $\mathbf{0.0172}$\\

ERGAS & & $5.6998$ & $4.7844$  & $3.2841$ & $3.8371$ & $3.5971$  & $4.8705$ & $5.8909$  & $2.1620$ & $4.1828$  & $3.2840$   & $\mathbf{2.1130}$   \\    

SAM & Chikusei & $9.5219$ & $7.9211$ & $3.9798$ & $6.0303$ & $4.9107$ & $8.0507$ &$8.5519$& $3.4601$ & $6.3220$  & $4.2936$ & $\mathbf{3.3667}$ \\

UIQI  & &  $0.8563$ & $0.8528$  & $0.9236$ & $0.9044$  & $0.9137$ & $0.8355$ & $0.7599$  & $0.9470$ & $0.9040$ & $0.9325$ & $\mathbf{0.9540}$ \\      
PSNR  & & $27.622$ & $27.472$ & $31.979$ & $28.861$ & $29.452$ & $27.185$ & $25.296$ & $34.742$  & $31.857$ & $34.057$ & $\mathbf{35.306}$  \\ \
SSIM && $0.9273$ & $0.9193$ & $0.9477$ & $0.9388$ & $0.9437$ & $0.8915$ & $0.8818$ & $0.9610$ & $0.8867$ & $0.9119$ & $\mathbf{0.9696}$\\ \hline
\hline
\end{tabularx}}
\end{adjustbox}
\caption{Performance comparison for Pavia and Chikusei datasets using standard quality metrics.}
\label{objectivemeasure_hsms}
\end{table*}


\begin{figure*}
\centering
\subfloat[Ground truth.]{\includegraphics[width=0.138\linewidth]{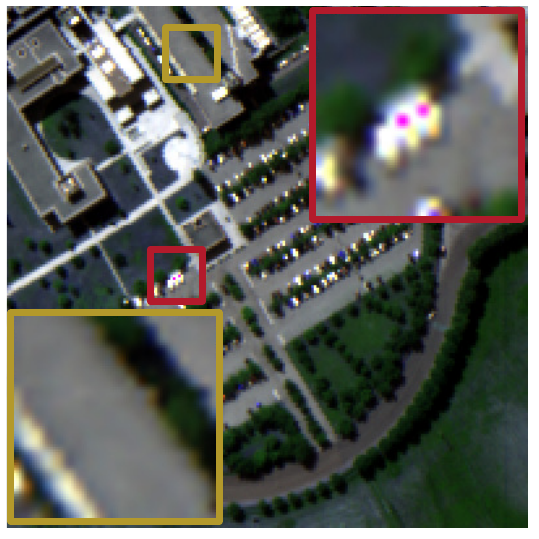}} \hspace{0.05mm}
\subfloat[Observed HS.]{\includegraphics[width=0.138\linewidth]{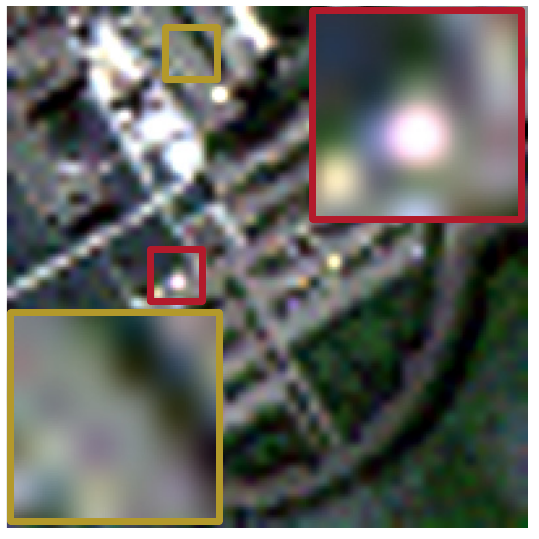}} \hspace{0.05mm}
\subfloat[Observed MS.]{\includegraphics[width=0.136\linewidth]{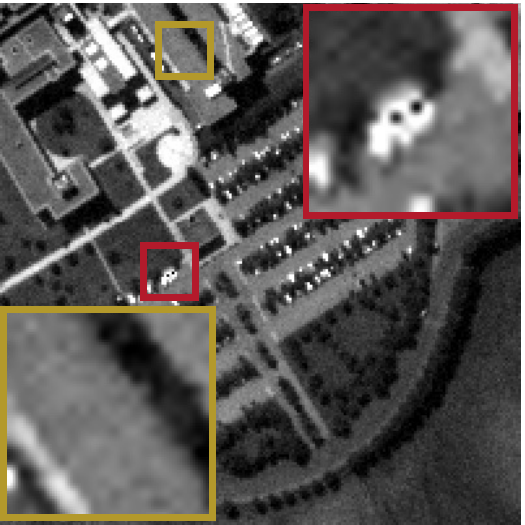}} \hspace{0.05mm}
\subfloat[GSA.]{\includegraphics[width=0.136\linewidth]{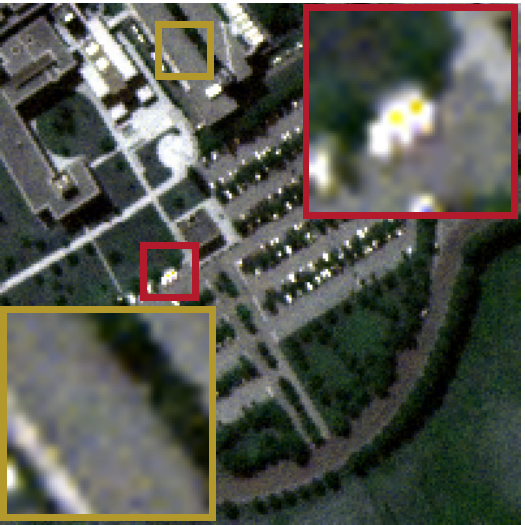}} \hspace{0.05mm}
\subfloat[CNMF.]{\includegraphics[width=0.136\linewidth]{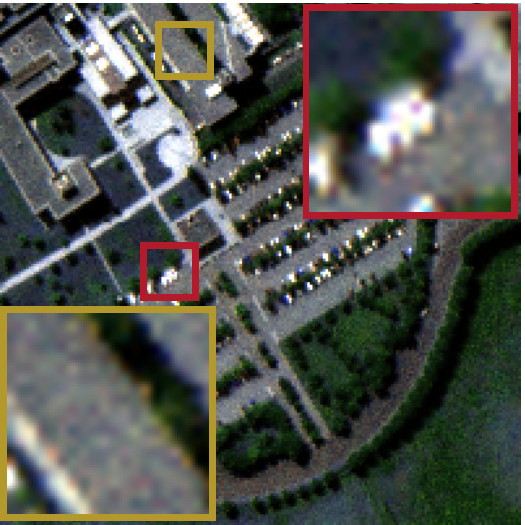}} \hspace{0.05mm}
\subfloat[JSU.]{\includegraphics[width=0.136\linewidth]{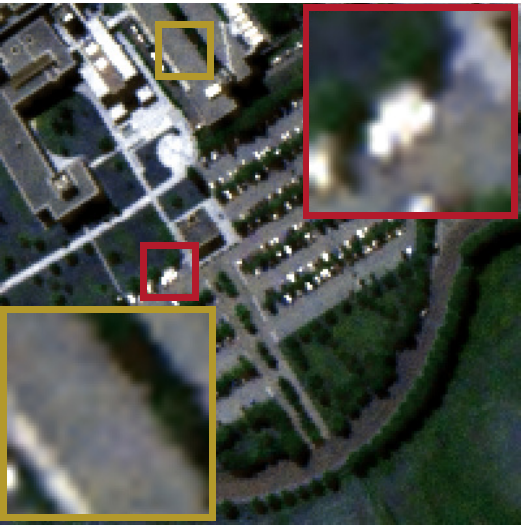}}  \hspace{0.05mm}
\subfloat[FUSE.]{\includegraphics[width=0.136\linewidth]{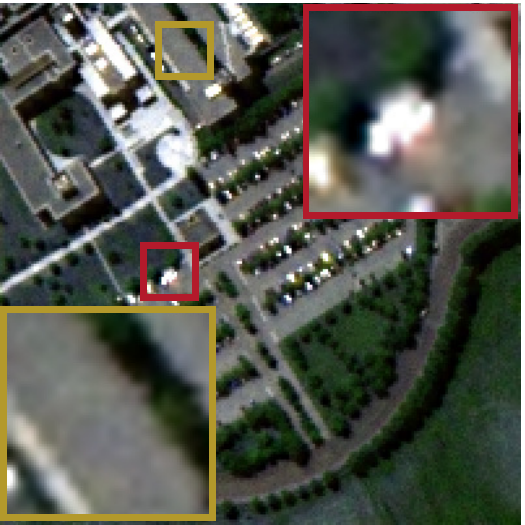}}  \hspace{0.05mm}\\
\subfloat[Sparse.]{\includegraphics[width=0.136\linewidth]{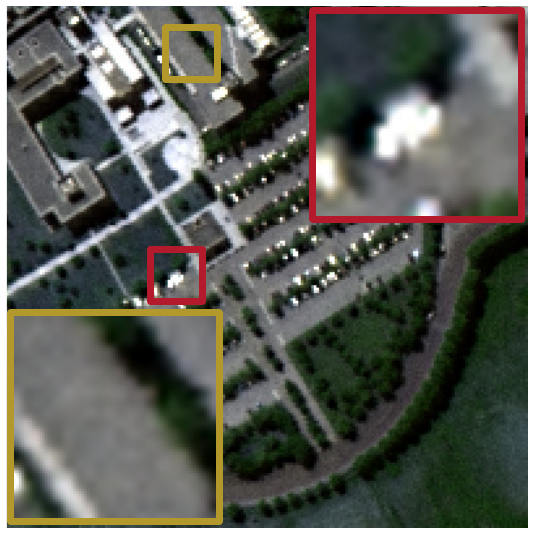}}  \hspace{0.05mm}
\subfloat[GLPHS.]{\includegraphics[width=0.136\linewidth]{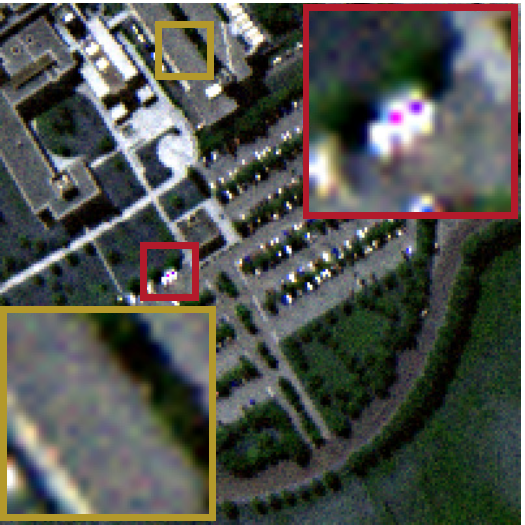}} \hspace{0.05mm}
\subfloat[MSMM.]{\includegraphics[width=0.136\linewidth]{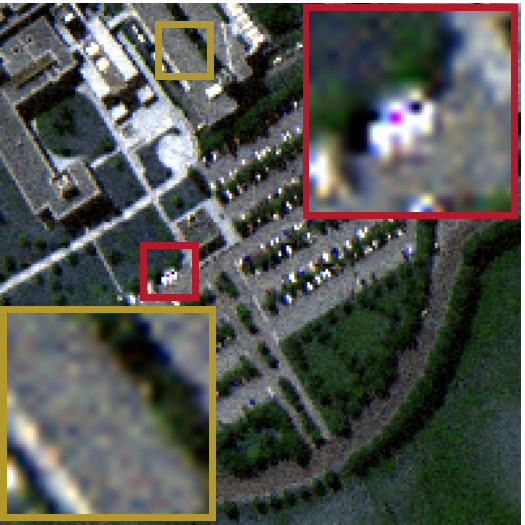}} \hspace{0.05mm}
\subfloat[NLSTF.]{\includegraphics[width=0.136\linewidth]{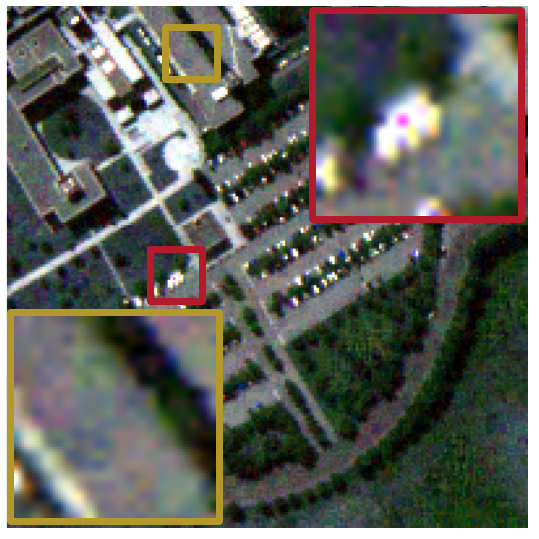}} \hspace{0.05mm}
\subfloat[SMBF.]{\includegraphics[width=0.136\linewidth]{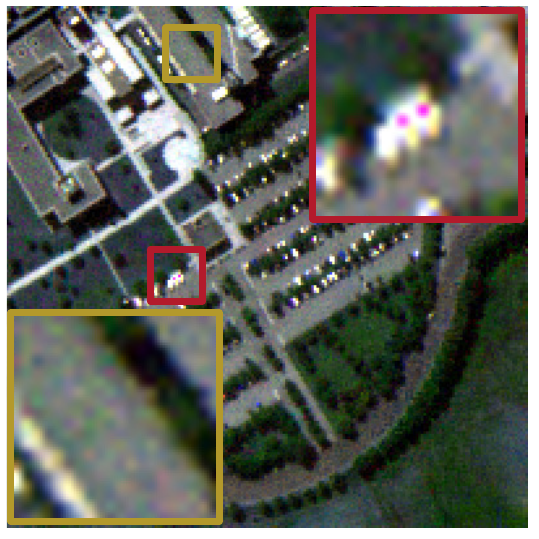}}  \hspace{0.05mm}
\subfloat[HySure.]{\includegraphics[width=0.136\linewidth]{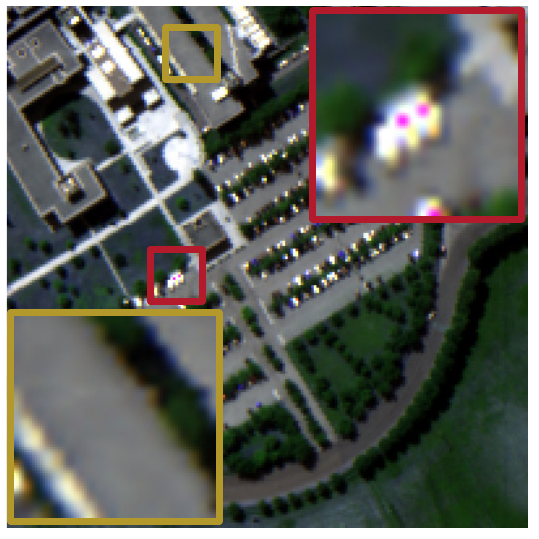}}  \hspace{0.05mm}
\subfloat[NLPR.]{\includegraphics[width=0.136\linewidth]{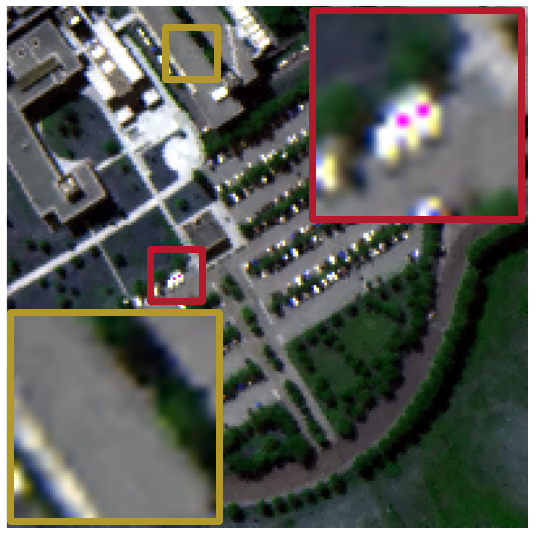}}  \hspace{0.05mm}
\caption{HS+MS fusion  for the Pavia dataset at $\text{SNR}_h$ and $\text{SNR}_\ell$ of $25$ dB.}
\label{pavia_results}
\end{figure*}

\begin{figure*}
\centering
\subfloat[Ground truth.]{\includegraphics[width=0.136\linewidth]{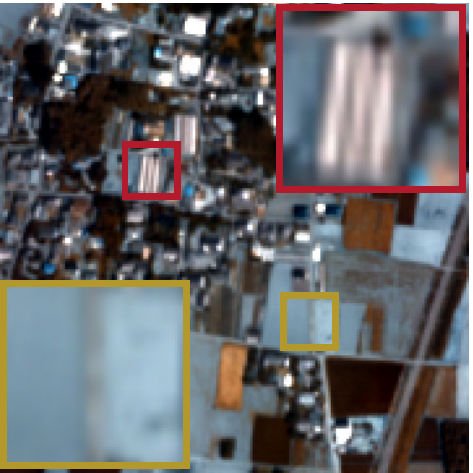}} \hspace{0.05mm}
\subfloat[Observed HS.]{\includegraphics[width=0.136\linewidth]{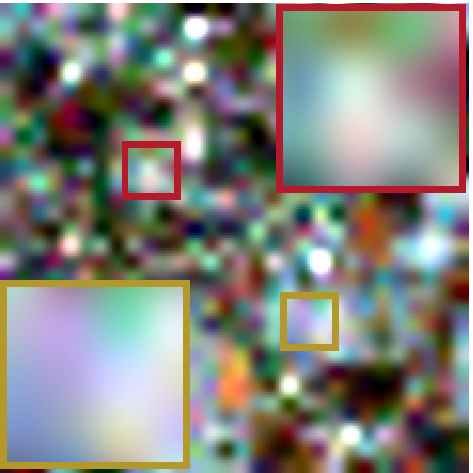}} \hspace{0.05mm}
\subfloat[Observed MS.]{\includegraphics[width=0.136\linewidth]{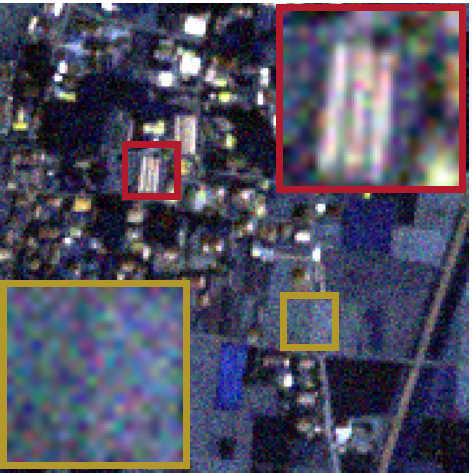}} \hspace{0.05mm}
\subfloat[GSA.]{\includegraphics[width=0.136\linewidth]{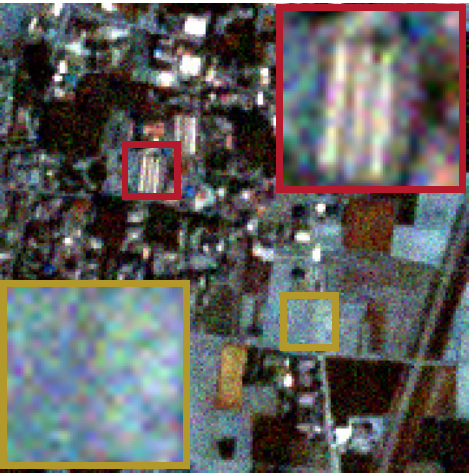}} \hspace{0.05mm}
\subfloat[CNMF.]{\includegraphics[width=0.136\linewidth]{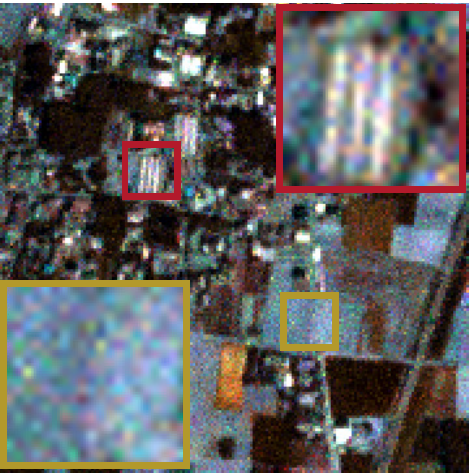}} \hspace{0.05mm}
\subfloat[JSU.]{\includegraphics[width=0.136\linewidth]{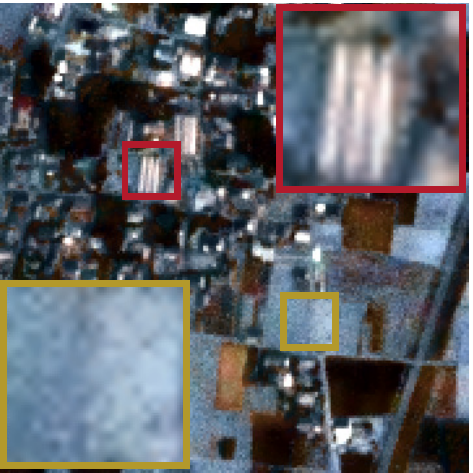}}  \hspace{0.05mm}
\subfloat[FUSE.]{\includegraphics[width=0.136\linewidth]{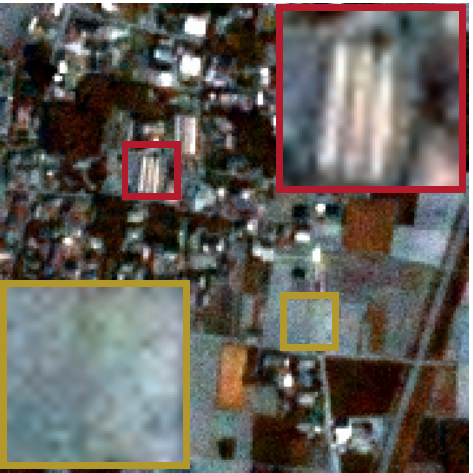}}  \hspace{0.05mm}\\
\subfloat[Sparse.]{\includegraphics[width=0.136\linewidth]{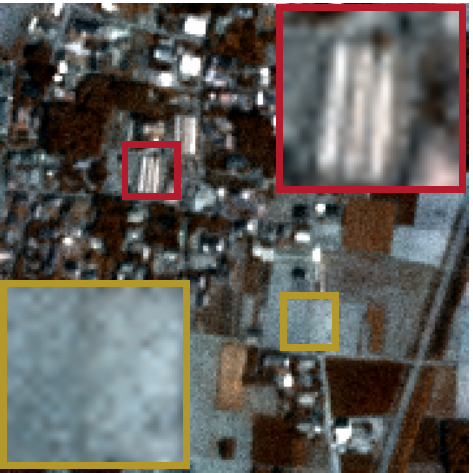}}  \hspace{0.05mm}
\subfloat[GLPHS.]{\includegraphics[width=0.136\linewidth]{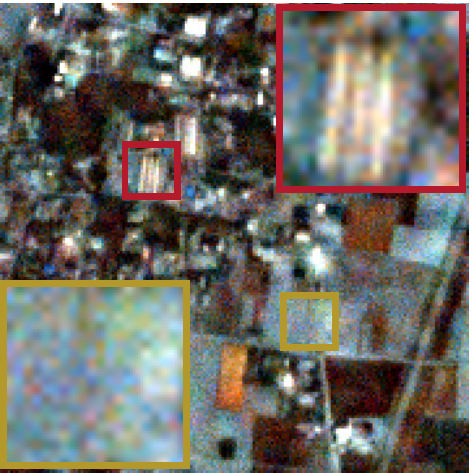}} \hspace{0.05mm}
\subfloat[MSMM.]{\includegraphics[width=0.136\linewidth]{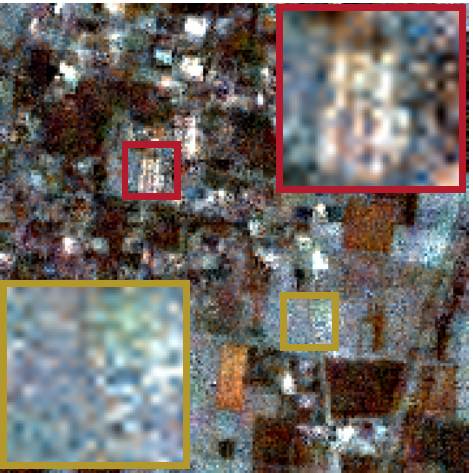}} \hspace{0.05mm}
\subfloat[NLSTF.]{\includegraphics[width=0.136\linewidth]{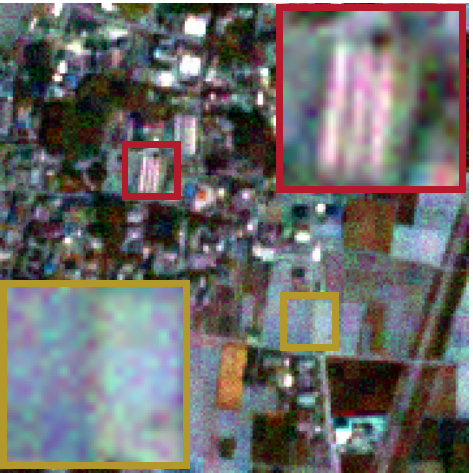}} \hspace{0.05mm}
\subfloat[SMBF.]{\includegraphics[width=0.136\linewidth]{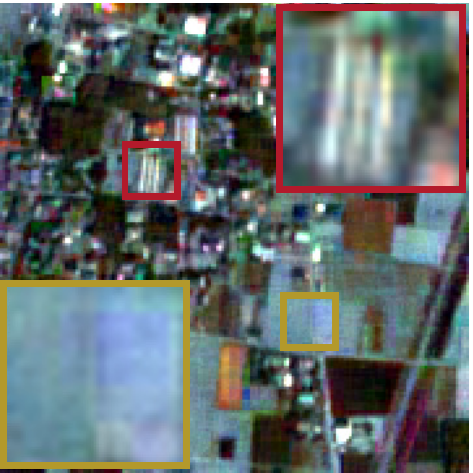}}  \hspace{0.05mm}
\subfloat[HySure.]{\includegraphics[width=0.136\linewidth]{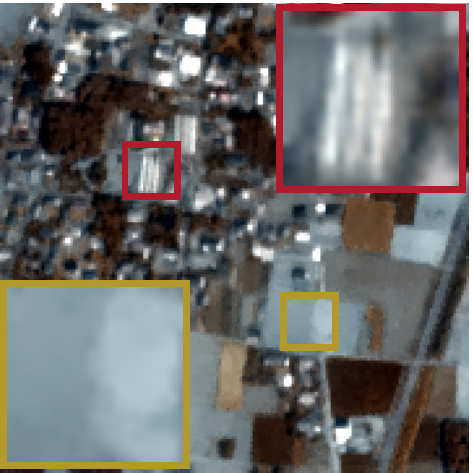}}  \hspace{0.05mm}
\subfloat[NLPR.]{\includegraphics[width=0.136\linewidth]{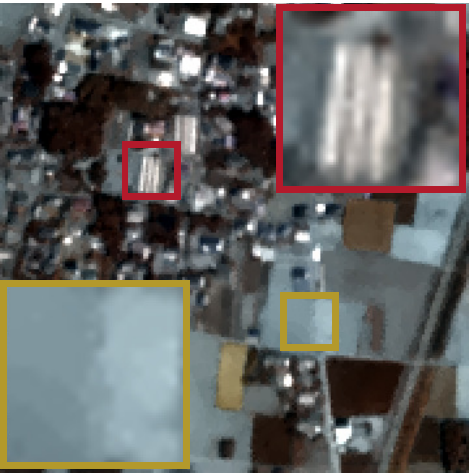}}  \hspace{0.05mm}
\caption{HS+MS fusion for the Chikusei dataset at $\text{SNR}_h$ and $\text{SNR}_\ell$ of $20$ dB.}
\label{chik_results}
\end{figure*}


\begin{figure*}[h]
\centering
\subfloat[Ground truth.]{\includegraphics[width=0.136\linewidth]{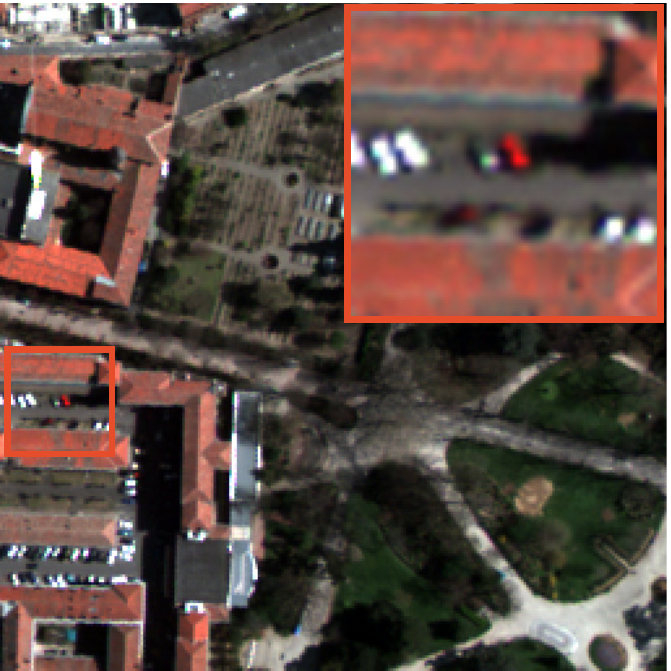}} \hspace{0.05mm}
\subfloat[Observed PAN.]{\includegraphics[width=0.136\linewidth]{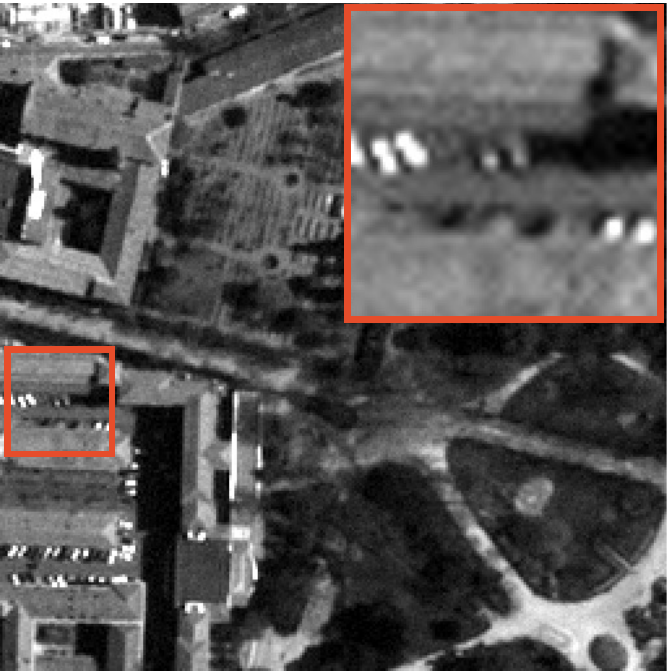}} \hspace{0.05mm}
\subfloat[Observed MS.]{\includegraphics[width=0.136\linewidth]{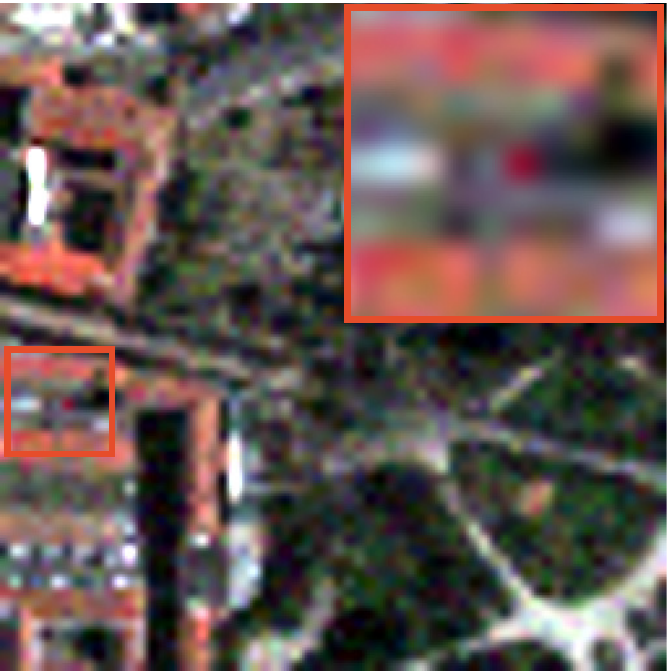}} \hspace{0.05mm}
\subfloat[GSA.]{\includegraphics[width=0.136\linewidth]{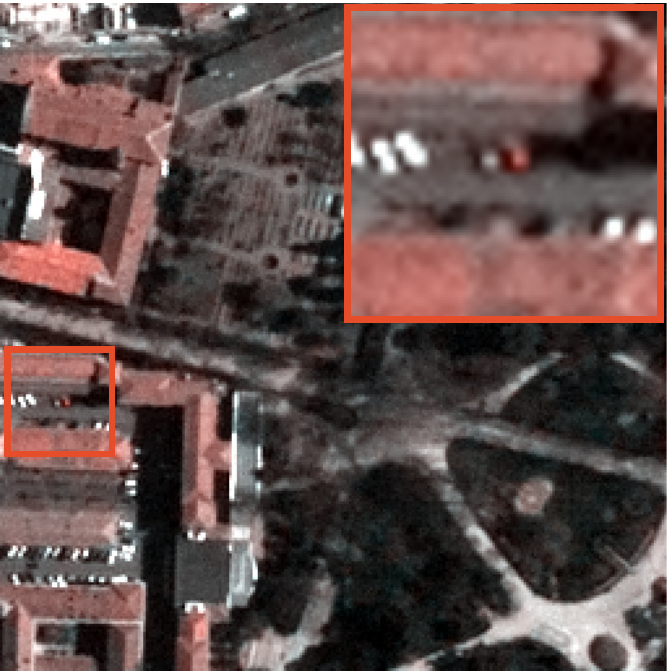}} \hspace{0.05mm}
\subfloat[FUSE.]{\includegraphics[width=0.136\linewidth]{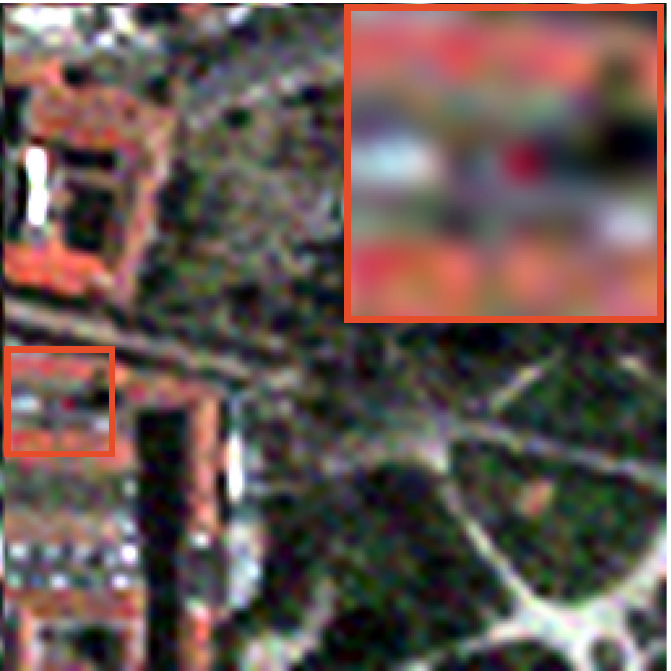}}  \hspace{0.05mm}
\subfloat[GLPHS.]{\includegraphics[width=0.136\linewidth]{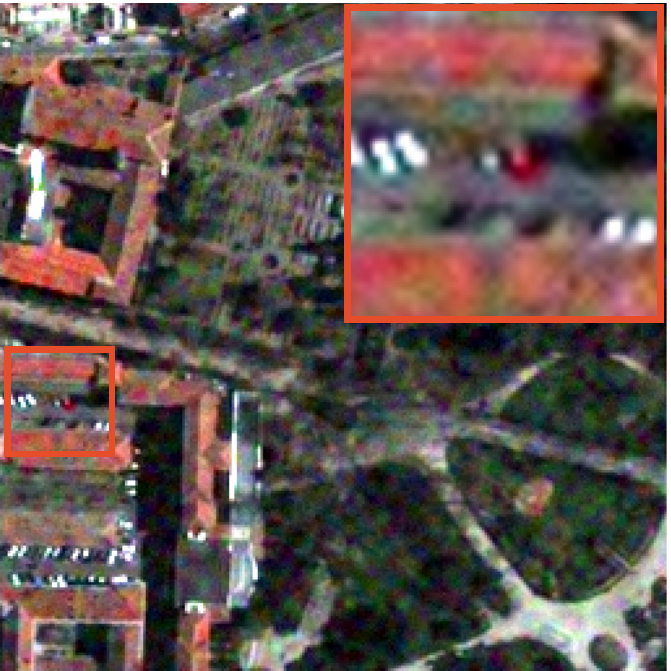}}  \hspace{0.05mm}
\subfloat[HySure.]{\includegraphics[width=0.136\linewidth]{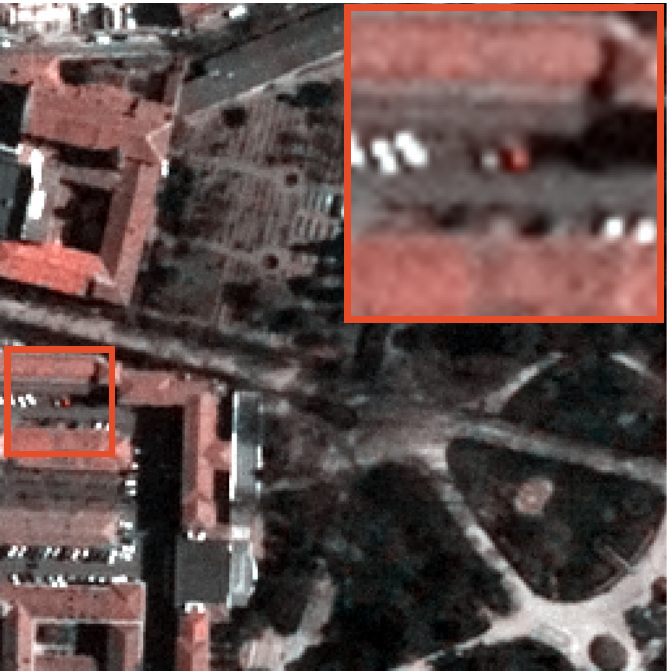}} \hspace{0.05mm}\\
\subfloat[BDSD.]{\includegraphics[width=0.136\linewidth]{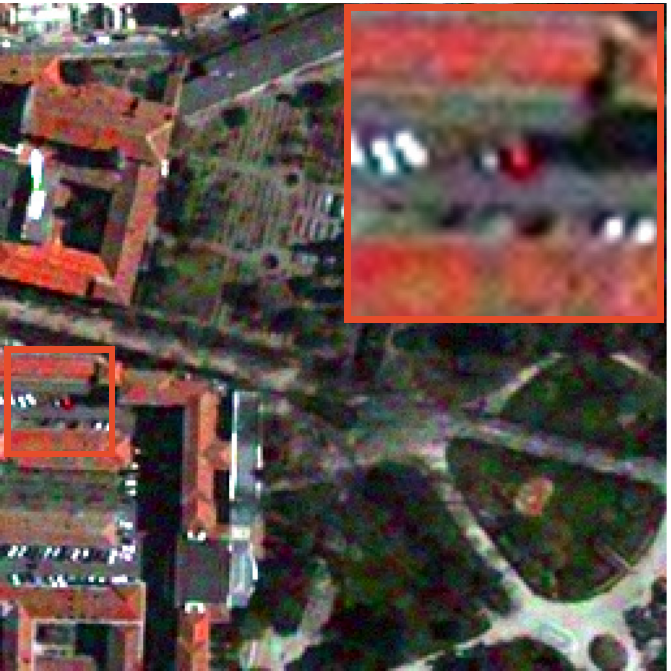}} \hspace{0.05mm}
\subfloat[PRACS.]{\includegraphics[width=0.136\linewidth]{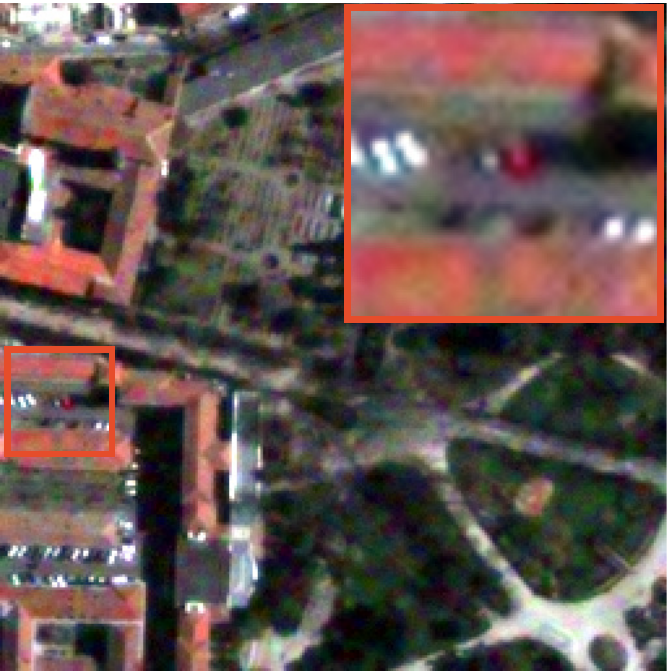}} \hspace{0.05mm}
\subfloat[SFIM.]{\includegraphics[width=0.136\linewidth]{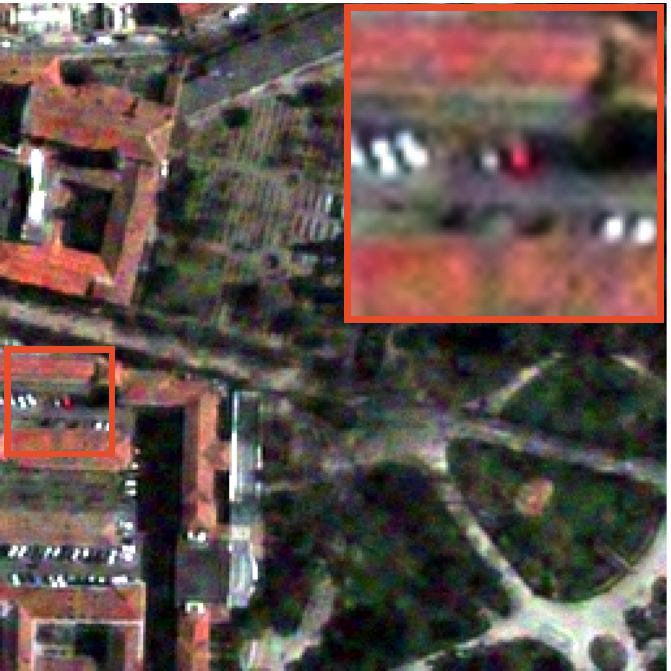}}  \hspace{0.05mm}
\subfloat[Indusion.]{\includegraphics[width=0.136\linewidth]{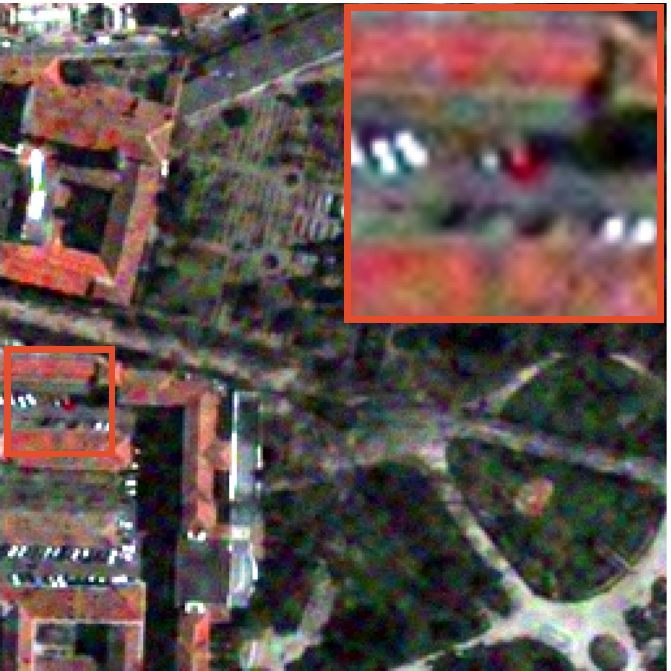}}  \hspace{0.05mm}
\subfloat[MGLP.]{\includegraphics[width=0.138\linewidth]{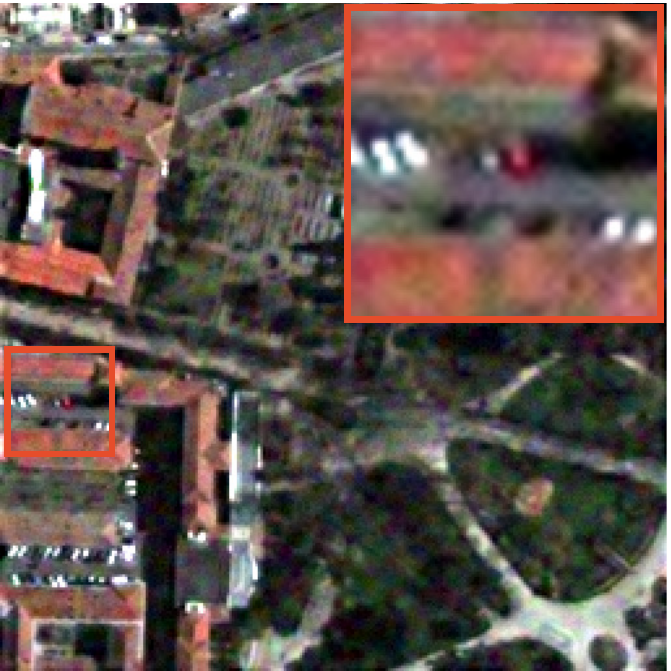}}  \hspace{0.05mm}
\subfloat[LGC.]{\includegraphics[width=0.138\linewidth]{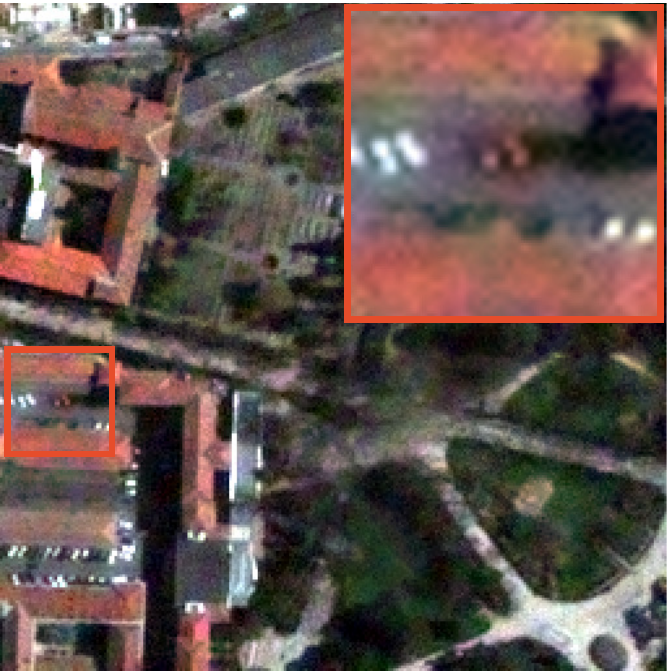}}  \hspace{0.05mm}
\subfloat[NLPR.]{\includegraphics[width=0.138\linewidth]{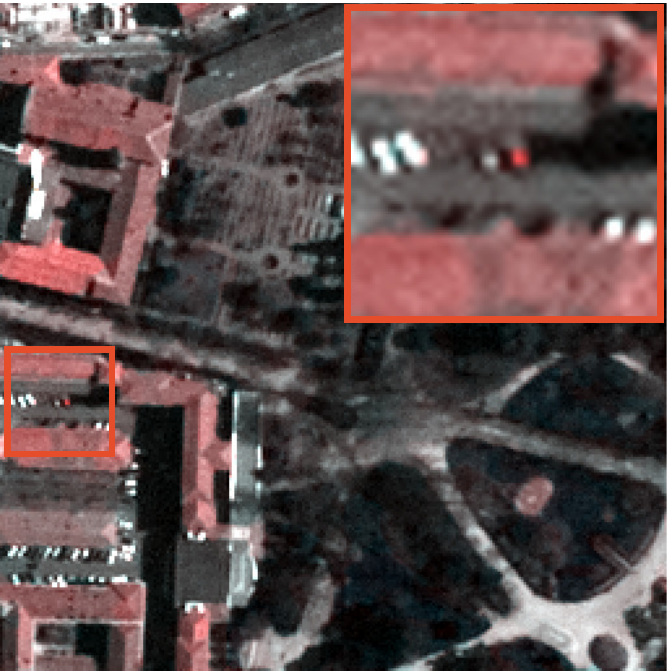}}  \hspace{0.05mm}
\caption{MS+PAN fusion for the Pl\'eiades dataset at $\text{SNR}_h$ and $\text{SNR}_\ell$ of $30$ dB and $20$ dB respectively.}
\label{mspan_results}
\end{figure*}

\begin{figure*}[h]
	\centering
	\subfloat{\includegraphics[width=0.245\linewidth]{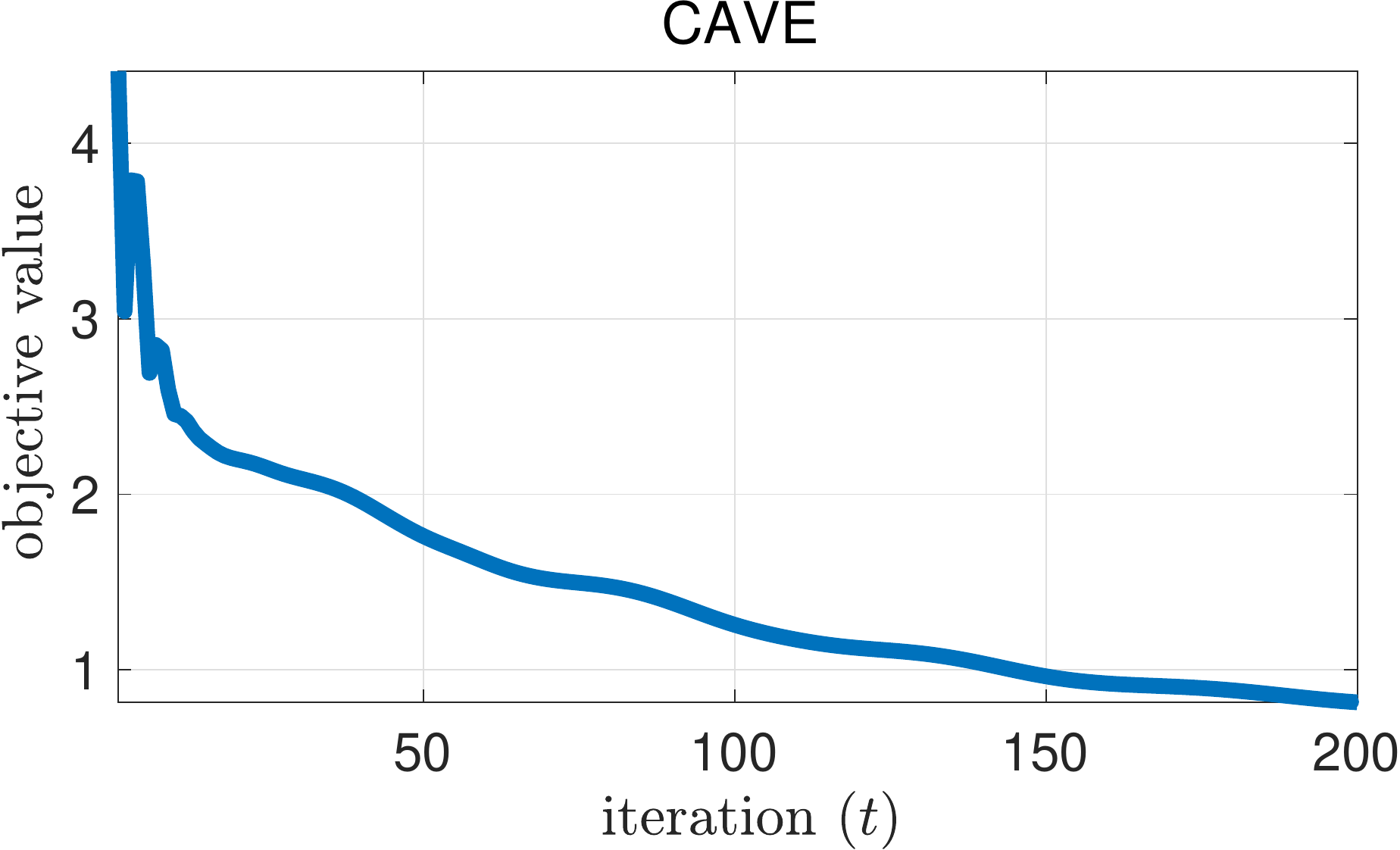}} \hspace{-0.55mm}
	\subfloat{\includegraphics[width=0.249\linewidth]{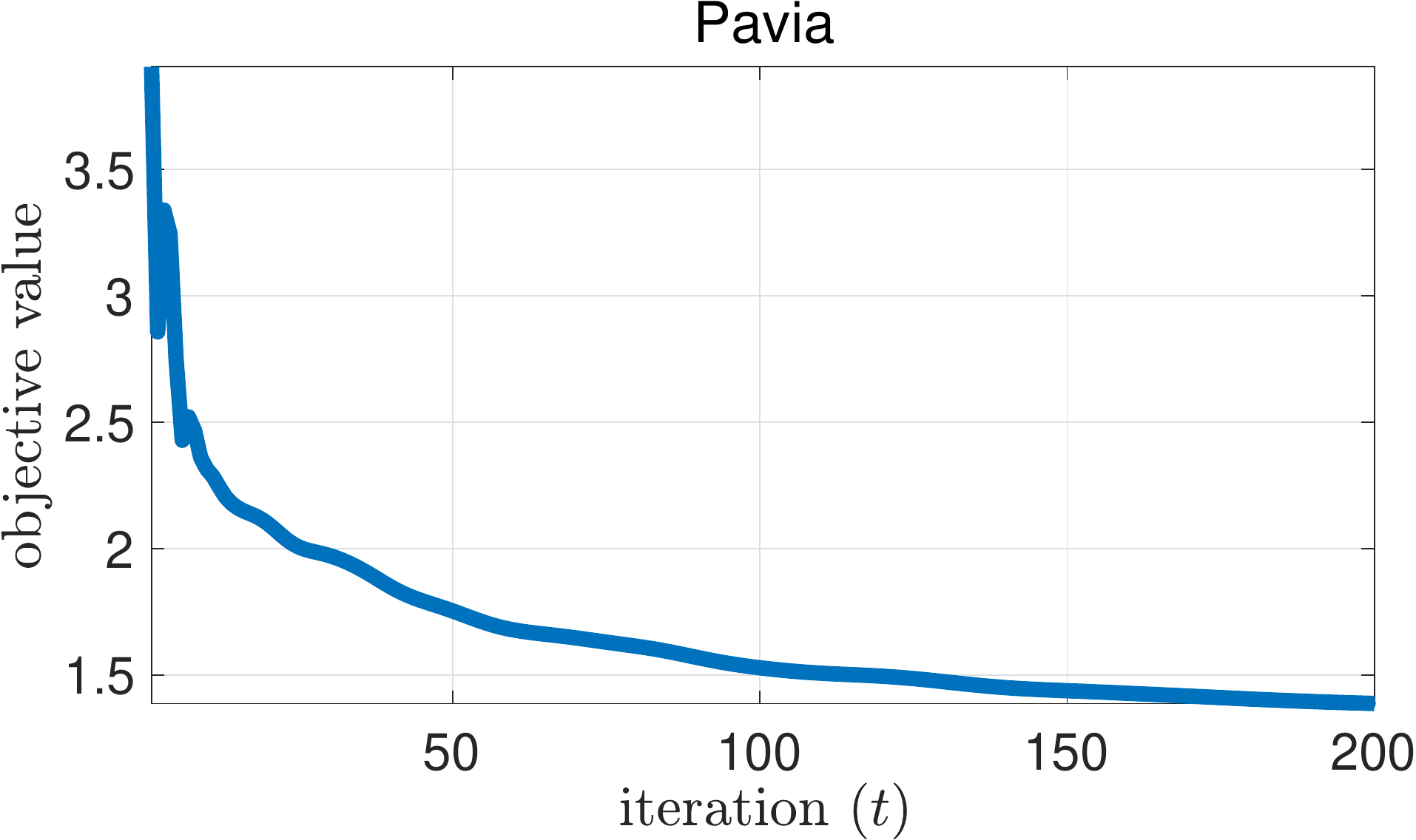}} \hspace{-0.55mm}
	\subfloat{\includegraphics[width=0.25\linewidth]{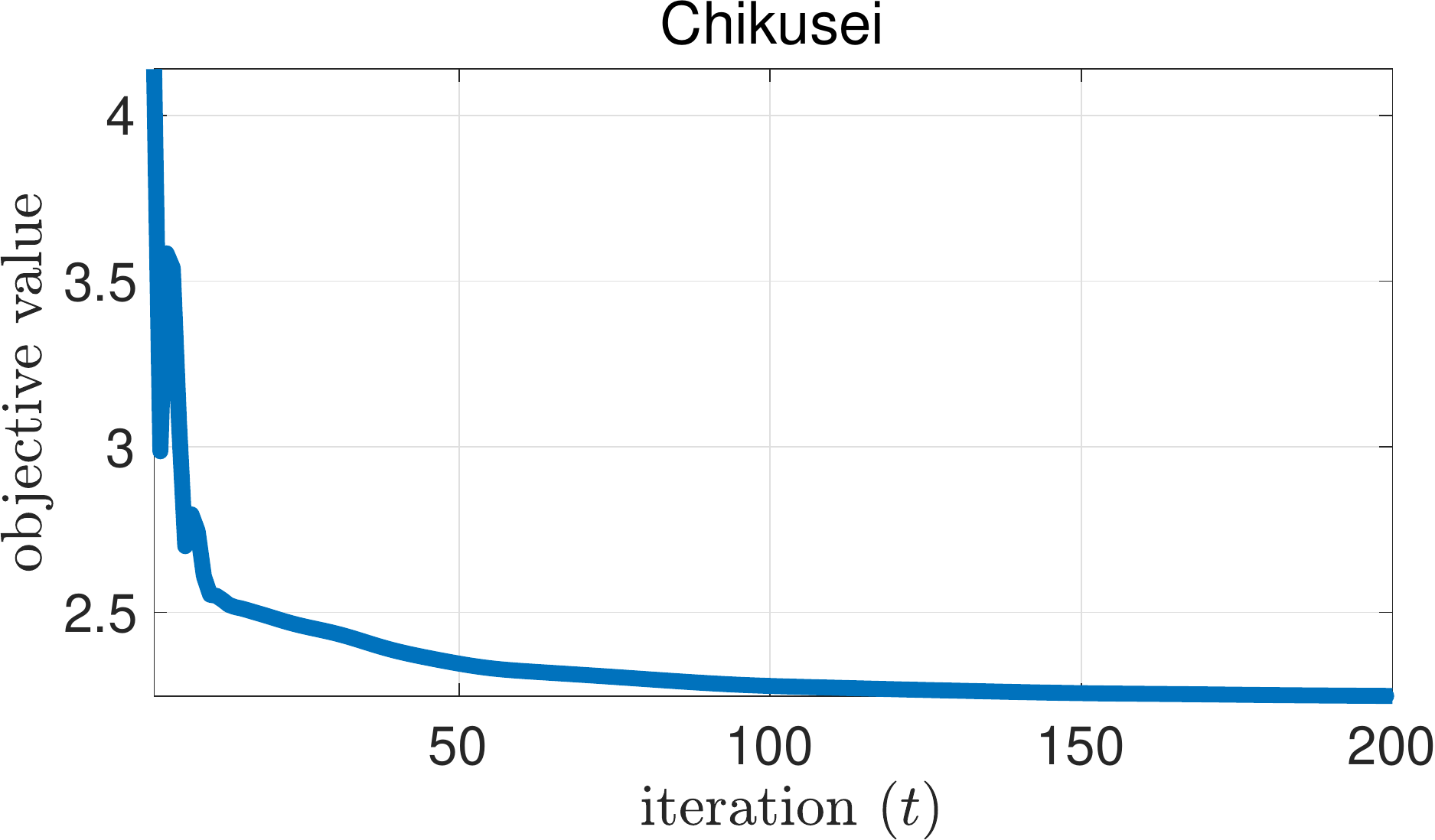}} \hspace{-0.55mm}
	\subfloat{\includegraphics[width=0.245\linewidth]{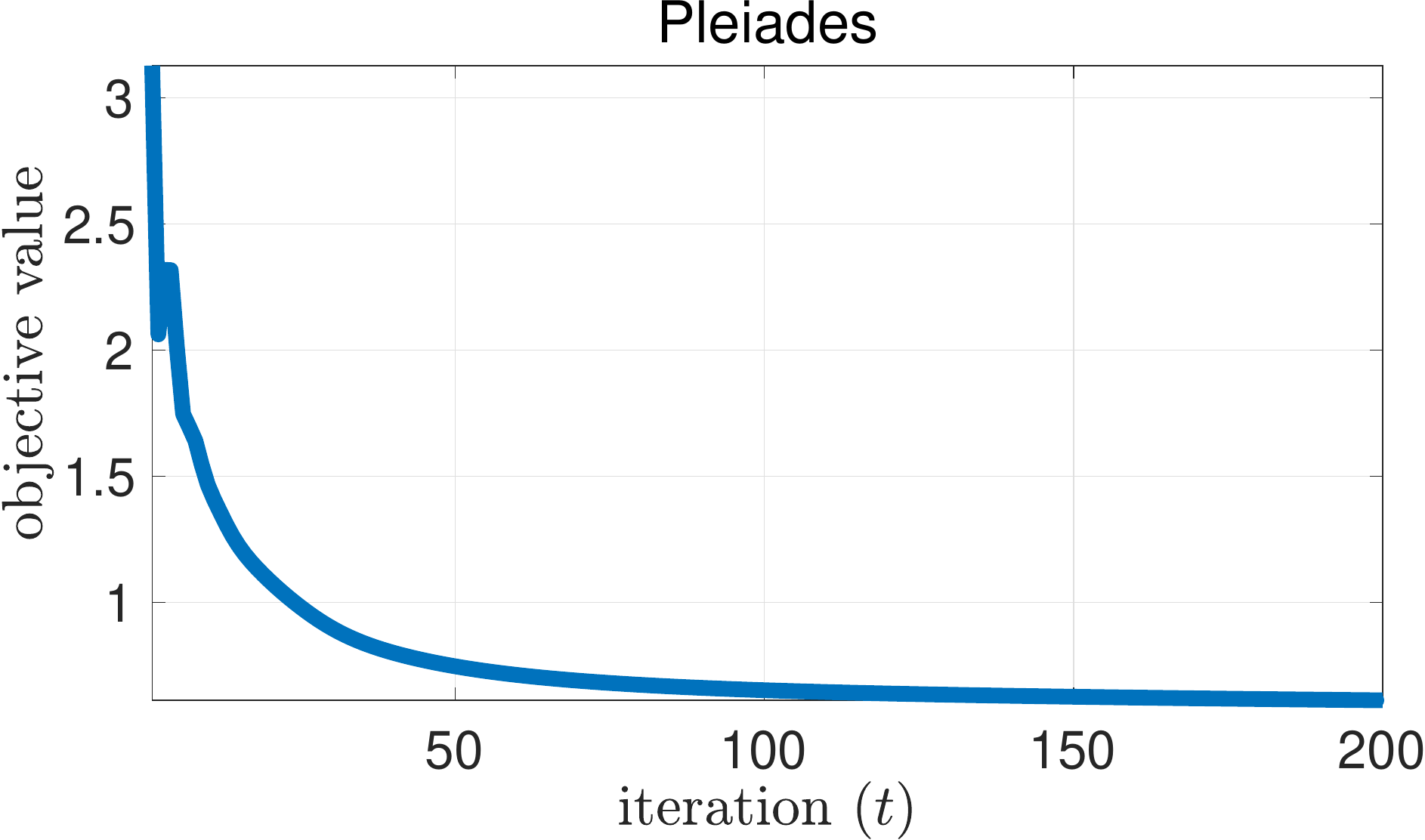}} \hspace{-0.55mm}
	\caption{Plot of objective value (log scale) as a function of iterations. CAVE, Pavia, and Chikusei are used for HS+MS fusion and Pl\'eiades  for Pansharpening.}
	\label{objective_conv}
\end{figure*}

\begin{table*}[!htp]
\centering
\scalebox{0.98}{
\begin{tabularx}{\textwidth}{c |*{12}{Y}}
\hline
Methods     & GSA    & LGC & FUSE   & GLPHS  & HySure & BDSD  & PRACS   & SFIM  & Indusion  &MGLP  & NLPR\\ 
\hline
\hline
RMSE&$0.0691$ &   $0.0279$ &   $0.0412$   & $0.0705$   & $0.0275$ &   $0.0333$   & $0.0289$  &  $0.0306$   & $0.0343$  &  $0.0320$ &   $\textbf{0.0268}$ \\
ERGAS   & $6.4510$   & $\textbf{2.8803}$  & $5.0371$   & $6.7092$  &  $3.0910$ &   $3.4071$   & $3.0683$   & $3.3834$  &  $3.5754$   & $3.4865$  &  $2.9212$ \\
SAM  &  $11.0215$ &   $3.9762$ &   $4.7341$  & $11.1728$&  $5.0311$   & $6.1943$  &  $4.3693$   & $4.3057$ &  $5.4619$  &  $4.6987$   & $\textbf{3.9636}$ \\
UIQI &   $0.7859$  &  $0.9309$ &   $0.8318$  &  $0.7682$ &  $0.9410$ &   $0.9136$  &  $0.9225$   & $0.9119$  &  $0.8955$  & $0.9157$  &  $\textbf{0.9411}$ \\
 PSNR & $23.310$ &  $31.116$  & $27.715$ &  $23.127$  & $31.273$  & $29.573$  & $30.810$  & $30.307$  & $29.320$   &$29.906$ &  $\textbf{31.429}$\\
SSIM& $0.8095$ &  $0.8884$  & $0.7769$ &  $0.7974$  & $\textbf{0.9139}$  & $0.8831$  & $0.8905$  & $0.8814$  & $0.8653$   &$0.8812$ &  $0.9122$
\\ \hline
\hline
\end{tabularx}}
\caption{Performance comparison for the Pl\'eiades dataset (averaged over $5$ images) using standard quality metrics.}
\label{objectivemeasure_mspan}
\end{table*}

\subsection{Database}

\begin{itemize}
\item CAVE: This dataset \cite{CAVE_0293} consists of 32 images, each of size $512 \times 512$ and $31$ spectral bands ($400-700$ nm).
We simulated the spectral response $\R$ following the protocol in \cite{cvpr11,xie2019multispectral} which was used to generate $\Y_h$; $\Y_\ell$ was obtained by $32\times$ downsampling of the ground-truth.

\item HS+MS: We considered two datasets. The Pavia dataset has $115$ spectral bands ($0.43-0.86$ $\mu$m) and a spatial resolution of $1.3$m. To simulate the $\Y_\ell$ image, we spatially blurred the ground-truth using the Starck–Murtagh filter \cite{starck1994image} and then downsampled the result by $4\times$ along both directions. To create the $\Y_h$ image ($4$ bands), the spectral response of IKONOS satellite was used. The Chikusei data set consists of HS images with $128$ bands ($0.36 - 1.018$  $\mu$m) and MS images with $9$ bands. The $\Y_\ell$ image is created similar to the Pavia dataset but with a $6\times$ downsampling. The spectral response $\R$ is estimated using the technique in \cite{simoes2015convex}. We conducted experiments on an HS image of size $50 \times 50 \times 93$ ($93$ bands and $50 \times 50$ resolution) and an MS image of size $200 \times 200 \times 4$ from the Pavia dataset. The size of the HS and MS images for Chikusei  are $30 \times 30 \times 128$ 
and $180 \times 180 \times 9$. Both $\text{SNR}_h$ and $\text{SNR}_\ell$ was set at $25$dB for Pavia  and at $20$dB for Chikusei.

\item MS+PAN: For pansharpening, we considered the Pl\'eiades dataset, where each MS image is of $60$ cm resolution and has four bands (RGB and NIR). The size of the panchromatic and MS images are $5000 \times 20000$ and $1250 \times 5000$. 
The spectral response was estimated as in \cite{simoes2015convex}, and the Starck–Murtaghh filter \cite{starck1994image} was used for the spatial blur.
We cropped the dataset into 5 non-overlapping regions of size $256 \times 256$, conducted the experiments on them individually, and averaged the results. 

\end{itemize}

\subsection{Compared methods}

We compared our algorithm with the following methods:
\begin{itemize}
	\item Component substitution: GSA \cite{aiazzi2007improving} and PRACS \cite{pracs}.
	\item Multiresolution analysis: GLPHS \cite{selva2015hyper},  SFIM \cite{sfim}, Indusion \cite{indusion}, and MGLP \cite{mtfglp}.
	\item Spectral unmixing: CNMF \cite{yokoya2012coupled} and JSU \cite{iccv15}.
	\item Variational methods: FUSE \cite{wei2015fast}, Sparse \cite{wei2015hyperspectral}, MSMM \cite{eismann2004application}, HySure \cite{simoes2015convex}, BDSD \cite{bdsd}, and LGC \cite{fu2019variational}.
	\item Tensor-based methods: NLSTF \cite{cvpr17} and SMBF \cite{TCYB20}.
	\item CNN-based methods: MHF \cite{xie2019multispectral}.
\end{itemize}


 \subsection{Quantitative metrics}
The spectral and spatial quality of the fused image is measured using five standard metrics \cite{wald2002data}---root mean squared error (RMSE), erreur relative globale adimensionnelle de synthese (ERGAS), spectral angle mapper (SAM), universal image quality index (UIQI), peak signal-to-noise ratio (PSNR) and structural similarity index measure (SSIM) \cite{wald2002data,yokoya2017hyperspectral}. Small values of ERGAS, SAM, and RMSE and large values of UIQI, PSNR, and SSIM signify better fusion quality.

\subsection{Parameter settings}

The various internal parameters of  our method and their settings are described below.
\begin{itemize}
\item Patch size and search size: We fixed the search and patch size to be $3 \times 3$ for all the experiments. A slight increase in performance is noted if we increase the search size to $5 \times 5$, but this also increases the cost.
\item Width $h$ in \eqref{wts}: We tuned this manually depending upon the noise level in the MS image; in general, $h$ is set proportionate to the noise level.
\item ADMM parameter $\rho$:  While theoretical convergence is guaranteed for any positive $\rho$ (Theorem \ref{maintheorem}), the convergence rate in practice is sensitive to $\rho$. For the datasets considered, we found that $\rho$ in the range $1\mathrm{e}\mbox{-}4$ to $0.1$ yielded the best results. We observed that smaller values of $\rho$ resulted in slower convergence.
\end{itemize}
The exact settings of $(\lambda_1,\lambda_2,\rho,h,L_s)$  for different experiments and datasets are as follows:
\begin{itemize}
\item Figure \ref{cave_results} (CAVE): $(0.7,1\mathrm{e}\mbox{-}4,1\mathrm{e}\mbox{-}3,0.15,8)$.
\item Figure \ref{pavia_results} (Pavia): $(0.8,2\mathrm{e}\mbox{-}4,1\mathrm{e}\mbox{-}3,0.15,20)$.
\item Figure \ref{chik_results} (Chikusei): $(1,1\mathrm{e}\mbox{-}3,0.095,0.25,20)$.
\item Figure \ref{mspan_results} (Pl\'eiades): $(0.85,9\mathrm{e}\mbox{-}3,1\mathrm{e}\mbox{-}3,0.17,4)$.
\end{itemize}

\subsection{Fusion results}

\textbf{\underline{CAVE}}: We compare the fusion quality visually and also after averaging the quantitative metrics on five random images from the CAVE dataset \cite{CAVE_0293} at $\text{SNR}_h$ and $\text{SNR}_\ell$ of $30$dB. The quality metrics for different methods are compared in table \ref{objectivemeasure_cave}. A sample fusion result  is shown in figure \ref{cave_results}, where we have only displayed the $16$th band. Our method shows less spectral distortion compared to competing methods  (compare the zoomed regions) and also has lesser noise grains. We compare our method with  a state-of-the-art deep learning method \cite{xie2019multispectral}, which has been shown to be superior to fusion networks such as 3D-CNN \cite{palsson2017multispectral} and PCNN \cite{scarpa2018target}.
Note that MHF-net is able to preserve spectral quality but shows noise; the proposed method is better able to retain spectral information and sharp edges and suppress noise. This is reflected in the lower ERGAS and SAM indices (table \ref{objectivemeasure_cave}).

\textbf{\underline{HS+MS}}:  The quality metrics for Pavia and Chikusei are shown in table \ref{objectivemeasure_hsms}. A visual comparison with the best performing methods is shown in figures \ref{pavia_results} and \ref{chik_results}. As before, we see that our spectral quality is better than competing methods  (compare the red dots in the zoomed regions in figures \ref{pavia_results} and \ref{chik_results}) and shows less noise. Although SMBF \cite{TCYB20} is able to retain the spectral quality, its performance is compromised in smooth regions. Also,  notice that the local pixel regularizer in HySure is unable to retain fine structures which are blurred out. Our regularizer is able to preserve fine structures with good spectral quality and this is reflected in the quality metrics. 

\textbf{\underline{MS+PAN}}: We perform the experiments at  $\text{SNR}_h$ and $\text{SNR}_\ell$  of $30$dB and $25$ dB respectively. The reconstructions are shown in figure \ref{mspan_results} and quality metrics in  table \ref{objectivemeasure_mspan}. Notice that the our method is able to retain fine structures. Although the competing methods were able to superresolve the MS image, the spectral degradation is noticeably large for them.  We compare our algorithm with a state-of-the-art variational method \cite{fu2019variational}, which is known to have superior performance over deep learning methods such as PNN \cite{scarpa2018target} and PAN-net \cite{yang2017pannet}. 
Notice that we are competitive with \cite{fu2019variational} though the latter method is customized for pansharpening.

\textbf{\underline{Comparison with nonlocal regularizers}}: Aside from the experiments above, we also compared our regularizer with existing nonlocal regularizers. The authors in \cite{duran2014nonlocal,duran2017survey} proposed a nonlocal regularizer for pansharpening where squared weighted $\ell_2$ norm (SWL2) of the nonlocal pixel gradients are penalized. Different from this, the authors in \cite{duran2018restoration} penalized the weighted $\ell_1$ norm (WL1) of the nonlocal gradients, while the weighted $\ell_2$ norm (WL2) is penalized in \cite{mifdal2021variational}. The quality metrics for the Pavia dataset are compared in Table \ref{nonlocal_comparison}. We see that SWL2 performs poorly. This is not a surprise, since  SWL2 can be seen as a nonlocal extension of the Sobolev prior \cite{peyre}; it assumes uniform smoothness of the image. WL1 and WL2 are better able to recover sharp features such as edges. Of these, WL1 which is based on the $\ell_1$ norm performs better. We see from  Table \ref{nonlocal_comparison} that the fusion results obtained using our regularizer are consistently better than WL1, WL2, and SWL2 in terms of quality metrics, especially for ERGAS and SAM.

\begin{table}
	\centering
	\scalebox{1.1}{
		\begin{tabularx}{0.42\textwidth}{c |*{8}{Y}}
			\hline
			
			Methods     & NLPR   &  SWL2   &  WL1  & WL2   \\ 
			
			\hline
			
			\hline
	RMSE &$\textbf{0.0125}$ &$0.0134$ &$0.0129$ &$0.0132$\\
	ERGAS&$\textbf{1.8432}$ &$2.0733$ &$1.9268$ &$2.0112$\\
	SAM  &$\textbf{2.8760}$ &$3.7834$ &$2.8809$ &$3.3412$\\
	UIQI &$\textbf{0.9879}$ &$0.9835$ &$0.9872$ &$0.9840$\\
	PSNR &$\textbf{38.039}$ &$37.461$ &$37.764$ & $37.560$\\
	SSIM &$\textbf{0.9710}$ &$0.9396$ &$0.9688$ &$0.9544$
			\\ \hline
			\hline
	\end{tabularx}}
	\caption{Comparison with different nonlocal regularizers on the Pavia dataset.}
	\label{nonlocal_comparison}
\end{table}

For completeness, we validate the convergence result in Theorem \ref{maintheorem} for the datasets used in figures \ref{cave_results}, \ref{pavia_results}, \ref{chik_results} and \ref{mspan_results}. In particular, we plot the objective function at each iteration. The results are shown in figure \ref{objective_conv}. We see that the objective stabilizes to the desired optimum within 200 iterations.

\begin{table}
	\centering
	\scalebox{1.1}{
		\begin{tabularx}{0.42\textwidth}{c |*{8}{Y}}
			\hline
			
			Methods     & $C_1$    &  $C_2$   &  $C_3$     &  $C_4$    &  $C_5$ \\ 
			
			\hline
			
			\hline
	RMSE &$\textbf{0.0125}$ &$0.0137$ &$0.0133$ &$0.0131$ &$0.0152$\\
	ERGAS&$\textbf{1.8432}$ &$2.0770$ &$2.0987$ &$2.0235$ &$2.5725$\\
	SAM  &$\textbf{2.8760}$ &$2.9927$ &$3.2223$ &$3.1998$ &$3.4064$\\
	UIQI &$\textbf{0.9879}$ &$0.9857$ &$0.9837$ &$0.9841$ &$0.9820$\\
	PSNR &$\textbf{38.039}$ &$37.217$ &$37.519$ &$37.684$ &$36.234$ \\
	SSIM &$\textbf{0.9710}$ &$0.9685$ &$0.9542$ &$0.9554$ &$0.9448$
			\\ \hline
			\hline
	\end{tabularx}}
	\caption{Ablation results on the Pavia dataset.}
	\label{ablation}
\end{table}

\subsection{Ablation study}

Finally, to understand the relative contribution of different components of our regularizer, we conduct an ablation analysis on the Pavia dataset.
Based on the choice between patch/pixel, weighted/unweighted and local/nonlocal, we consider the following five important cases:
\begin{itemize}
	\item $C_1$: Nonlocal weighted patch regularization in which we implement $\phi(\X)$ as  in \eqref{reg1}, with weights $\omega_{\i\btau}$ given  by \eqref{wts}.
	\item  $C_2$:  Nonlocal unweighted patch regularization in which we implement   $\phi(\X)$ as in \eqref{reg1}, but with weights $\omega_{\i\btau}$ set to 1.
	\item$C_3$: Nonlocal weighted pixel regularization in which the weights are computed using (4) but pixels are used instead of patches for regularization. Note that the weights are computed as per (4) using patches from the MS image.

\item $C_4$: Nonlocal unweighted pixel regularization, where the weights $\omega_{\i \btau}$ are set to 1, and pixels are used instead of patches for regularization.

\item $C_5$: Local unweighted pixel-based regularizer, where the weights $\omega_{\i \btau}$ are set to 1 and we consider only local pixel differences for regularization.
\end{itemize}
The results are shown in Table \ref{ablation}. We notice that the quality metrics are highest for the case $C_1$ where both patches and nonlocal weighting are used (proposed regularizer). On the other hand, the quality metrics are lowest for $C_5$ that uses local pixels and no weighting. We found that weighting and patches have an equal impact on the fusion quality---similar performance is obtained for the case $C_2$ (where patches are used but the weighting is turned off) and the case $C_3$ (where pixels are used along with weighting). The former is seen to exhibit slightly better performance in terms of spectral quality metrics such as ERGAS and SAM. Case $C_4$ (where pixels are used and weighting is turned off) performs worse than $C_1$, $C_2$ and $C_3$, but performs better than its local counterpart $C_5$.



\section{Conclusion}
\label{conclusion}
In this paper, we proposed a variational multiband image fusion technique using a nonlocal patch-based regularizer. We showed that the regularizer is more effective than local pixel-based regularizers and helps in preserving self-similar structures such as textures in remote sensing images.  We also came up with a novel means of encoding patch variations using convolutions, and showed how (along with an appropriate variable splitting technique) this can be used to derive an ADMM algorithm where the subproblems have closed-form solutions. We presented exhaustive results on HS-MS fusion and pansharpening which show that our regularizer is competitive with state-of-the-art variational and deep learning methods and is better able to capture fine textures in remote sensing images.

\section{Appendix: Proof of Theorem 1}
\label{admm}

Convergence of ADMM for convex programs is well established \cite{eckstein1992douglas,boyd2011distributed,bauschke2011convex,teodoro2019convergent}. However, convergence typically holds under some technical conditions, which have to be verified for the optimization problem at hand. Unlike standard optimization problems where the variable is a vector, the optimization variable in \eqref{ouroptim} is a matrix. However, both vectors and matrices can be seen as elements of some finite-dimensional inner-product space and existing convergence results can be extended in a straightforward manner to this more general setting \cite{bauschke2011convex}. In particular, as explained next, the ADMM convergence results in  \cite{teodoro2019convergent} can be applied to our problem.

Let $\mathcal{H}_1$ and $\mathcal{H}_2$ be two finite dimensional real vector spaces and let $\mathcal{A}: \mathcal{H}_1 \to \mathcal{H}_2$ be a linear operator. Consider the optimization problem
\begin{equation}
\label{teoderooptim}
\underset{\X  \in \mathcal{H}_1}{\min} \ f\big(\X\big) + g\big(\mathcal{A}(\X)\big),
\end{equation}
where $f: \mathcal{H}_1 \to \Re$ and $g: \mathcal{H}_2 \to \Re$ are closed convex functions. 

An ADMM algorithm for \eqref{teoderooptim} is as follows: 
\begin{align}
&\X^{(t+1)} = \underset{\X}{\mathrm{min}} \ f(\X) +  \frac{\rho}{2} \| \mathcal{A}(\X) - \V^{(t)} - \U^{(t)} \|_\ast \label{xupdate}\\
&\V^{(t+1)} = \underset{\V}{\mathrm{min}} \ g(\V) +  \frac{\rho}{2} \| \mathcal{A}(\X^{(t+1)}) - \V - \U^{(t)} \|_\ast \label{vupdate}\\
&\U^{(t+1)} = \U^{(t)} + \mathcal{A}(\X^{(t+1)}) - \V^{(t+1)}, \label{uupdate}
\end{align}
where $\X \in \mathcal{H}_1$ and $\V \in \mathcal{H}_2$ are the primal variables, $\U \in \mathcal{H}_2$ is the dual variable and the norm used in the updates is induced by some arbitrary inner product $\inner{\cdot}{\cdot}_\ast$  defined on $\mathcal{H}_2$. 
By adapting the following convergence result in \cite{teodoro2019convergent}, one can show that the above ADMM algorithm solves \eqref{teoderooptim}.
\begin{theorem}
\label{maintheorem}
Assume that the problem \eqref{teoderooptim} is solvable, i.e., there exists $\X^*$ where the minimum is attained. Also, assume that the linear operator $\mathcal{A}$ is injective. Then, for any arbitrary initialization and any $\rho>0$, the sequence  $(\X^{(t)})_{t \geqslant 0}$ generated using \eqref{xupdate}, \eqref{vupdate} and \eqref{uupdate} converges to $\X^*$.
\end{theorem}

We note that our optimization problem \eqref{ouroptim} is an instance of \eqref{teoderooptim}, where 
\begin{itemize}
\item  $\mathcal{H}_1 = \Re^{n_m \times L_s}$ and $\mathcal{H}_2 = \Re^{n_m (2 + |W||P|) \times L_s}$, with the Frobenius inner product on $\mathcal{H}_1$ and $\mathcal{H}_2$. 
\item  $\mathcal{A}$ is defined as
\begin{equation}
\label{Vdef}
\mathcal{A}(\X) = 
\begin{pmatrix}  
\B\X \\ 
\X \\ 
\W
\end{pmatrix},
\end{equation}
where $\W \in \Re^{n_m |W||P| \times L_s}$ is obtained by stacking the matrices $\{\D_{\btau\k}\X \in \Re^{n_m \times L_s}: \btau \in W, \k \in P\}$ along rows. Clearly, $\mathcal{A}$ is injective since $\mathcal{A}(\X)=\mathbf{0}$ implies $\X=\mathbf{0}$. 
\item $f$ is the zero function and $g: \mathcal{H}_2  \to \R$ is defined as
\begin{align}
\label{gfunction}
\hspace*{-6cm}&\quad g(\V) = \frac{1}{2}{\|\Y_\ell\!-\!\S\V_1\E\|}^2\!+\!\frac{\lambda_1}{2}{\|\Y_h\!-\!\V_2\E\R\|}^2 \nonumber \\
&+ \frac{\lambda_2}{2}\sum_{\i \in \Omega} \sum_{\btau \in W} \sum_{\k \in P} \sum_{c=1}^{L_s}  \omega_{\i\btau} | \V_{\btau\k}(i,c)|, 
\end{align} 
where
\begin{equation}
\label{Vdef2}
\V = 
\begin{pmatrix}  
\V_1\\ 
\V_2 \\ 
\V_3
\end{pmatrix}
\end{equation}
where $\V_1$, $\V_2 \in \Re^{n_m \times L_s}$ and $\V_3 \in \Re^{n_m |W||P| \times L_s}$ is the row-wise concatenation of $|W||P|$ matrices $\{\V_{\btau\k} \in \Re^{n_m \times L_s}: \btau \in W, \k \in P\}$.
\end{itemize}

Next we need to set up  the correspondence between updates \eqref{x_update_problem}, \eqref{P1_update_problem}, \eqref{P2_update_problem}, \eqref{Qj_update_problem}, \eqref{dualvar1_main}, \eqref{dualvar2_main}, \eqref{dualvar3_main} of the proposed ADMM algorithm and the updates in \eqref{xupdate}-\eqref{uupdate}. The primal variables $\X$ and $\V$ in \eqref{xupdate}-\eqref{uupdate} and the primal variables in \eqref{x_update_problem}-\eqref{dualvar3_main} can be related as: $\V_1 = \P_1, \V_2 = \P_2,$ and $\V_{\btau\k} = \Q_{\btau\k}$. The dual variables can be related as 
\begin{equation}
\label{Udef}
\U = 
\begin{pmatrix}  
\BLambda_1 \\ 
\BLambda_2 \\ 
\BSigma
\end{pmatrix}
\end{equation}
where $\BSigma \in \Re^{n_m |W||P| \times L_s}$ is a matrix obtained by the row-wise concatenation of $\{\BSigma_{\btau\k} \in \Re^{n_m \times L_s}: \btau \in W, \k \in P\}$. 

The correspondence between updates \eqref{xupdate}-\eqref{uupdate} and \eqref{x_update_problem}-\eqref{dualvar3_main} is now clear: Update \eqref{xupdate} corresponds to \eqref{x_linear_system_main}; update  \eqref{vupdate} corresponds to  \eqref{Qh_update}, \eqref{Qm_update} and \eqref{Qj_main}, since $\V$ is separable in $\P_1,\P_2$ and $\{\Q_{\btau\k}\}$; and update \eqref{uupdate} corresponds to \eqref{dualvar1_main}, \eqref{dualvar2_main} and \eqref{dualvar3_main}, since $\U$ is separable in $\BLambda_1,\BLambda_2$ and $\{\BSigma_{\btau\k}\}$.  Convergence of the proposed ADMM algorithm now follows from Theorem \ref{maintheorem}.

{\small
\bibliographystyle{IEEEbib}
\bibliography{egbib}
}

\end{document}